\documentclass[useAMS,usenatbib]{mn2e}
\def\Msun {\,\mathrm{M}_\odot}

\usepackage{amsmath}
\usepackage{amssymb}
\usepackage{graphicx}
\usepackage{color}
\usepackage{url}
\title[Mono-enriched second generation stars]{Descendants of the first stars: the distinct chemical signature of second generation stars}

\author[T. Hartwig et al.]
{\parbox{\textwidth}{
Tilman Hartwig$^{1,2,3,4}$\thanks{E-mail: tilman.hartwig@ipmu.jp}, Naoki Yoshida$^{1,2}$,  Mattis Magg$^{5}$, Anna Frebel$^6$, Simon C. O. Glover$^{5}$, Facundo A. G{\'o}mez$^{7,8}$, Brendan Griffen$^6$, Miho N. Ishigaki$^2$, Alexander P. Ji$^{9,16}$, Ralf S. Klessen$^{5,10}$, Brian W. O'Shea$^{11,12,13,14}$, Nozomu Tominaga$^{2,15}$}\\
\vspace{2mm}\\
$^1$Department of Physics, School of Science, University of Tokyo, Bunkyo, Tokyo 113-0033, Japan\\
$^2$Kavli IPMU (WPI), The University of Tokyo, Kashiwa, Chiba 277-8583, Japan\\
$^3$Sorbonne Universit\'es, UPMC Univ Paris 06, UMR 7095, Institut d'Astrophysique de Paris, 75014 Paris, France\\
$^4$CNRS, UMR 7095, Institut d'Astrophysique de Paris, 75014, Paris, France\\
$^5$Universit\"at Heidelberg, Zentrum f\"ur Astronomie, ITA, Albert-Ueberle-Stra{\ss}e 2, 69120 Heidelberg, Germany\\
$^6$Department of Physics and Kavli Institute for Astrophysics and Space Research, MIT, Cambridge, MA 02139, USA\\
$^{7}$Instituto de Investigaci{\'o}n Multidisciplinar en Ciencia y
Tecnolog{\'i}a, Universidad de La Serena, Ra{\'u}l Bitr{\'a}n 1305, La
Serena, Chile\\
$^{8}$Departamento de F{\'i}sica y Astronom{\'i}a, Universidad de La
Serena, Av. Juan Cisternas 1200 N, La Serena, Chile\\
$^{9}$The Observatories of the Carnegie Institution of Washington, 813 Santa Barbara St., Pasadena, CA 91101, USA\\
$^{10}$Universit\"at Heidelberg, Interdisziplin\"ares Zentrum f\"ur Wissenschaftliches Rechnen, INF 205, 69120 Heidelberg, Germany\\
$^{11}${Department of Computational Mathematics, Science and Engineering, Michigan State University, MI, 48823, USA}\\
$^{12}${Department of Physics and Astronomy, Michigan State University, MI, 48823, USA}\\
$^{13}${National Superconducting Cyclotron Laboratory, Michigan State University, MI, 48823, USA}\\
$^{14}${Joint Institute for Nuclear Astrophysics - Center for the Evolution of the Elements, USA}\\
$^{15}$Department of Physics, Faculty of Science and Engineering, Konan University, 8-9-1 Okamoto, Kobe, Hyogo 658-8501, Japan\\
$^{16}$Hubble Fellow
}

\begin{document}

%\date{Accepted date. Received date; in original form date}

\pagerange{\pageref{firstpage}--\pageref{lastpage}} \pubyear{2017}

\maketitle

\label{firstpage}

\begin{abstract}
Extremely metal-poor (EMP) stars in the Milky Way (MW) allow us to infer the properties of their progenitors by comparing their chemical composition to the metal yields of the first supernovae. This method is most powerful when applied to mono-enriched stars, i.e. stars that formed from gas that was enriched by only one previous supernova. We present a novel diagnostic to identify this subclass of EMP stars.
We model the first generations of star formation semi-analytically, based on dark matter halo merger trees that yield MW-like halos at the present day. Radiative and chemical feedback are included self-consistently and we trace all elements up to zinc.
Mono-enriched stars account for only $\sim 1\%$ of second generation stars in our fiducial model and we provide an analytical formula for this probability.
We also present a novel analytical diagnostic to identify mono-enriched stars, based on the metal yields of the first supernovae. 
This new diagnostic allows us to derive our main results independently from the specific assumptions made regarding Pop~III star formation, and we apply it to a set of observed EMP stars to demonstrate its strengths and limitations.
Our results may provide selection criteria for current and future surveys and therefore contribute to a deeper understanding of EMP stars and their progenitors.
\end{abstract}


\section{Introduction}
%tbd: include new UMP star \citep{aguado18}.
The first stars in the Universe (the so-called ``Pop~III'' stars) are of fundamental importance for understanding galaxy formation. They enriched the primordial interstellar medium (ISM) and intergalactic medium with heavy elements, they contributed to the reionization of the Universe, and they played a crucial role in the formation of the first supermassive black holes. Owing to the lack of efficient coolants in metal-free gas, we expect the first stars to have a higher characteristic mass than is found for present-day star formation. Direct observations of the first stars to test the theories of their formation are also lacking. Our knowledge about the mass distribution of the first stars is thus mainly based on theoretical models and simulations \citep{glover13,greif15}. Another independent constraint is the absence of any low-mass Pop~III survivors in the Milky Way (MW), which limits the masses of the first stars to $\gtrsim 0.8\Msun$ \citep{bond81,hartwig15,komiya16b,ishiyama16,dutta17,magg18}.

Stellar archaeology provides a powerful approach to constrain the nature and properties of the first stars \citep{frebel15}. Spectroscopic observations of extremely metal-poor (EMP) stars in the MW enable measurements of their chemical composition. The relative abundances of the different elements can then be compared with the theoretically predicted yields of their putative progenitor supernovae (SNe). Several studies have successfully interpreted the abundance signatures of individual EMP stars as the fingerprint of Pop~III SNe, and obtained estimates for the stellar mass of the corresponding progenitor \citep{ishigaki14,tominaga14,keller14,ji15,placco15,placco16,fraser17,chen17,ishigaki18}. However, a major assumption of this reverse-engineering problem is that the EMP star carries the chemical imprint of only one SN. Accounting for metal contributions from several SNe would require additional free parameters, and consequently weakens the constraints due to degeneracies between the individual yields. %study with two SN fitted?.

A key challenge of stellar archaeology is therefore to identify mono-enriched second generation stars, as they are most valuable for constraining the properties of the first stars. Here, we define ``mono-enriched'' second generation stars as stars that formed from gas that was enriched by exactly one Pop~III SN. In contrast, we refer to stars that carry the combined chemical signature of more than one SN as ``multi-enriched''.

%\citet{ryan96} proposed that stars with a metallicity of [Fe/H]$<-2.7$ have formed after only one previous enrichment event.
%For the relative abundance of two elements A and B we use the standard notation

Metallicity alone is not a reliable tracer of the stellar population because the metallicity of gas enriched by a single Pop III SN depends sensitively on the metal yield of the SN, which varies greatly, particularly for an element such as Fe, and on the degree of metal mixing, which is poorly constrained. For example, in our models, we find mono-enriched second generation stars with metallicities\footnote{Defined as $[\mathrm{A}/\mathrm{B}] = \log_{10}(m_\mathrm{A}/m_\mathrm{B})-\log_{10}(m_{\mathrm{A},\odot}/m_{\mathrm{B},\odot})$, where $m_\mathrm{A}$ and $m_\mathrm{B}$ are the abundances of elements A and B and $m_{\mathrm{A},\odot}$ and $m_{\mathrm{B},\odot}$ are the solar abundances of these elements \citep{asplund09}.} $[{\rm Fe}/{\rm H}] > -3$ and later generations of star formation with metallicities as small as $[{\rm Fe}/{\rm H}] \sim -3$. The carbon-enhancement of most EMP stars has been claimed as an additional signature of second generation stars, emerging from faint Pop~III SNe \citep{beers92,aoki07,ishigaki14,asa15}. In this paper, we investigate further indicators and diagnostic to successfully identify mono-enriched second generation stars, based on their chemical abundance. This allows us to construct samples of stars that are mono-enriched based on our current understanding of Pop~III SNe.

A special subclass of second generation stars are those that form from gas that was enriched by a pair-instability supernova (PISN). These very energetic explosions of massive metal-poor stars are the final fates of non-rotating Pop~III stars in the mass range $140-260\Msun$ \citep{rakavy67,barkat67,fraley68,bond84,fryer01}. They eject more metals than core collapse SNe and can therefore enrich the ISM of their host halo to higher metallicities, beyond [Fe/H]$\sim -3$. This makes it more difficult to search for second generations stars that form from the debris of a PISN because the number of ordinary stars increases with metallicity and the fraction of PISN-enriched stars at [Fe/H]$> -3$ is very small \citep{deB17}. The nucleosynthetic yield of a PISN has a strong deficiency of the odd-charged elements \citep{heger02}, but this signature has not yet been conclusively observed in stellar archaeology surveys \citep{aoki14}. 
It is therefore crucial to derive the distinct chemical signature of second generation stars that form from gas enriched by a PISN. In this paper, we provide guidance to identify mono-enriched stars from core collapse or pair-instability SNe and also derive the completeness fraction of current stellar archaeology surveys that focus on [Fe/H]$<-3$.

%\citet{bovino14}: Formation of carbon-enhanced metal-poor stars in the presence of far ultraviolet radiation (simulation). 

%Previous works have focused on individual aspects of the formation of the second generation of stars.
%Semi-analytical models have studies the chemical evolution of the MW \citep{deB17,graziani17}. Simulations:  \citet{jeon17} (We perform a suite of cosmological hydrodynamic zoom-in simulations to follow their evolution from the era of the first generation of stars down to $z = 0$)

%Galactic/stellar archaeology, near-field cosmology.
%metal mixing (include in main text where appropriate): \citet{madau01,wise08,whalen08,greif07,greif10,jeon14,ss14,oshea15}.
%\citet{kobayashi14}: ``We show that the low ratios of $\alpha$ elements (Mg, Si, and Ca) to Fe recently found for a small fraction of extremely metal-poor stars can be naturally explained with the nucleosynthesis yields of core-collapse supernovae or hypernovae''. Hypernovae produce more iron than normal SNe \citep{nomoto03}

\section{Methodology}

\subsection{Semi-analytical model of star formation}
%Structure in our Universe forms hierarchically
Cosmic structure formation proceeds hierarchically from small matter overdensities in the early Universe via accretion and mergers. Hierarchical structure formation is dominated by dark matter, which accounts for most of the matter in the Universe. To model the baryonic physics of star and galaxy formation, we can therefore decouple the formation and mergers of dark matter halos and the physics and stellar feedback within them.

Our semi-analytical approach is based on dark matter merger trees that were separately generated from high-resolution $N$-body simulations. On top of this dark matter framework, we model star formation and the corresponding feedback self-consistently with a set of analytical recipes. For this study, we use 30 MW-like merger trees from the Caterpillar project \citep{griffen16}, which assumes the current dark energy plus cold dark matter ($\Lambda$CDM) paradigm with cosmological parameters from \citep{planck14}. The halos were selected based on three criteria to resemble the MW: virial masses in the range $0.7\times 10^{12} < M_\mathrm{vir}/\Msun < 3\times 10^{12}$, no other halos with $M_\mathrm{vir} > 7 \times 10^{13}\Msun$ within 7\,Mpc, and no halos with $M_\mathrm{vir}>0.5M_\mathrm{host}$ within $2.8$\,Mpc. The mass of a dark matter particle in the highest resolution zoom region is $3 \times 10^4\Msun$, which is sufficient to resolve also the smallest Pop~III star-forming halos at high redshift \citep{griffen17,magg18}. The time between snapshots at high redshift is ${\sim}5$\,Myr at $z > 6$ and ${\sim}50$\,Myr at $z < 6$. This guarantees a high temporal resolution to model accurately the radiative and chemical feedback of Pop~III stars. Our semi-analytical model of Pop~III star formation is based on \citet{hartwig15} with improvements by \citet{magg16,magg18}. For further details on the model and a resolution study see \citet{magg18}.

In the early Universe the main components of primordial gas clouds are hydrogen and helium with H$_2$ being the most efficient coolant under the conditions considered here. Once a pristine halo reaches the critical mass
\begin{equation}
\label{eq:mcrit}
M_\mathrm{crit} = 3.3 \times 10^6\Msun \left( \frac{1+z}{10} \right)^{-3/2},
\end{equation}
cooling by molecular hydrogen is efficient enough to allow the gas to collapse to protostellar densities and trigger star formation \citep{yoshida03, hummel12}. Massive stars forming in these halos produce large numbers of soft ultraviolet photons in the Lyman and Werner bands of H$_{2}$. These Lyman-Werner (LW) photons can readily escape from low-mass halos \citep{schauer15} and so the onset of Pop III star formation is quickly followed by the growth of an extragalactic LW background. We model the effect of this LW feedback as a uniform background that increases with time according to
\begin{equation}
F_{21}(z)= 4\pi 10^{-(z-10)/5},
\end{equation}
where $F_{21}$ has the units $10^{-21}\mathrm{erg}\,\mathrm{s}^{-1}\,\mathrm{cm}^{-2}\,\mathrm{Hz}^{-1}$ \citep{greif06}. Most halos are illuminated by a LW flux that is within a factor of two of this mean value \citep{dijkstra08}, which justifies this approximate treatment. LW photons can photodissociate H$_2$ and hence destroy the most important coolant in the early Universe and consequently prevent star formation. In addition to the critical mass required for primordial star formation (Eq. \ref{eq:mcrit}) we therefore check that the halo mass is above \citep{OShea08}
\begin{equation}
M_\mathrm{LW} = 5 \times 10^5\Msun + 3.5 \times 10^6 \Msun F_{21}^{0.47}.
\label{eq:LW}
\end{equation}
Baryonic streaming velocities might further alter this threshold and require a higher critical mass, but the relative importance of this effect is still debated (\citealt{stacy11,greif11,naoz13,tanaka14,schauer17,hirano17}; Schauer et al., in prep.).

Once we identify a halo in which Pop~III stars can form, we assign individual metal-free stars to it by sampling stochastically from a logarithmically flat initial mass function (IMF) until the total stellar mass is above
\begin{equation}
\label{eq:Mstar}
M_* = \eta _\mathrm{III} \frac{\Omega _b}{\Omega _m} M_\mathrm{h},
\end{equation}
where $\eta _\mathrm{III}$ is the star formation efficiency (SFE) of Pop~III stars and $M _\mathrm{h}$ is the mass of the halo. The SFE and the lower and upper limit of the Pop~III IMF are calibrated to match observational constraints (see Sec. \ref{sec:calib}). We assume that star formation is instantaneous and model the ionizing feedback on subsequent star formation. The emerging H{\sc ii} regions around star-forming halos suppress star formation in their vicinity by photoionization heating and we allow star formation in halos that are within the H{\sc ii} region of a neighbouring halo only if $T_\mathrm{vir}>10^4$\,K.%(for more details see \citealt{magg18}).

Once a star explodes as an SN, we follow the expansion of its metal-enriched shell. For Pop~III SNe we assume a constant velocity of 10\,km\,s$^{-1}$ in the intergalactic medium \citep{smith15} and for metals from SNe of later generation stars we model their expansion as a momentum-driven snowplough (see \citealt{magg18} for details on the ionizing feedback and external enrichment).

%these three paragraphs all started with "once..."
When a halo has been enriched with metals, the second generation of stars form from this enriched interstellar medium \citep[e.g.,][]{chiaki16}. We distinguish two different enrichment channels: if the halos has been enriched internally by Pop~III stars in the same halo, we delay the formation of second generation stars by the recovery time $t_\mathrm{recov}=100$\,Myr \citep{greif10,whalen13,smith15,jeon14,jeon17,chiaki18}. If a previously pristine halo is externally enriched and has a mass above $M_\mathrm{LW}$, Pop~II star formation occurs one freefall time after this enrichment with
\begin{equation}
t_\mathrm{ff} = 72\,\mathrm{Myr} \left( \frac{1+z}{10} \right)^{-3/2},
\end{equation}
where we assume an overdensity of 200 times the mean cosmic density.
%During this time, between the enrichment event and the formation of second generation stars, the host halo can merge with other metal-enriched halos or experience further enrichments, which are also taken into account.
In this paper, we refer to second generation stars as those that form after the first metal enrichment of a halo. Due to the delay between the first enrichment and the onset of second generation star formation, the host galaxy can be enriched by multiple enrichment events or merge with an already enriched galaxy before the second generation of stars forms.

The main topic of this paper are the nature, chemical characteristics, and unique signature of second generation stars. We assume that the composition of such a second generation star is defined at the moment of its formation and does not change during the lifetime due to possible pollution by ISM accretion (\citealt{tanaka17}, see also \citealt{yoshii81,frebel09,komiya10,hattori14,komiya15,johnson15,shen16}). Whenever we refer to the chemical composition of second generation stars, we implicitly refer to the chemical composition of the ISM from which these second generation stars form.%since we do not explicitly model the formation of individual second generation stars.

\subsection{SN yields and chemical enrichment}
\label{sec:yields}
One novel feature of our semi-analytical model is the tracking of chemical elements up to zinc. This enables us to calibrate our model based on various observations and we obtain crucial insight into the chemical enrichment history of the MW. In this section, we briefly summarize the main features of our model of chemical evolution.

For Pop~III stars, we use the tabulated metal yields as a function of progenitor mass by \citet{nomoto13}. The theoretical uncertainty for the metal yields between different models \citep{hw10,limongi12} is of the order $0.3$\,dex for carbon to zinc, as we will discuss below. Independent of the SN progenitor mass, we assume that $20\%$ of the ejected metals fall back within the recovery time and $80\%$ are ejected from the gravitational potential of the halo \citep{wise08,ritter12}. For internal and external enrichment, we assume instantaneous mixing and if more than one SN contributes to the enrichment, the individual metal yields are added. To model inhomogeneous mixing of the metals with the ISM, we assume that only a fraction $f_\mathrm{dil}$ of all hydrogen in the halo mixes with the metals. This approach is consistent with more advanced theoretical models \citep{starkenburg13,hirai17,chen17,sarmento17,sarmento17b} and we draw the dilution factors from a log-normal distribution with mean $\mu=10^{-1.5}$ and width $\sigma=0.75$\,dex. More realistic hydrodynamical simulations of the mixing of the first SN yields have been performed self-consistently in 3D by other groups \citep{greif07,wise08,whalen08,greif10,wise12,ritter12,vasiliev12,jeon14,ss14,ritter15,ritter16,sharma16,smith15,oshea15,chen17b}.

We do not account for metal enrichment by Type Ia SNe or red giant branch stars because these processes are expected to occur at later cosmic times and do not significantly contribute to the enrichment of second generation stars \citep{komiya16}.

To model the metal yields from Pop~II stars, we assume that $5\%$ of the stellar mass is eventually ejected as metals \citep{vincenzo16}. Since we are interested in the first enrichment events, we presume for simplicity that all of these metals are ejected by Type II SNe. To determine how this mass of metals is distributed over the individual elements, we apply the distribution of chemical yields by \citet{nomoto13} for stars at $Z=0.001$ and average the contribution by SNe with different progenitor masses over a Salpeter IMF in the range $10-40\Msun$.

\begin{figure}
\centering
\includegraphics[width=0.47\textwidth]{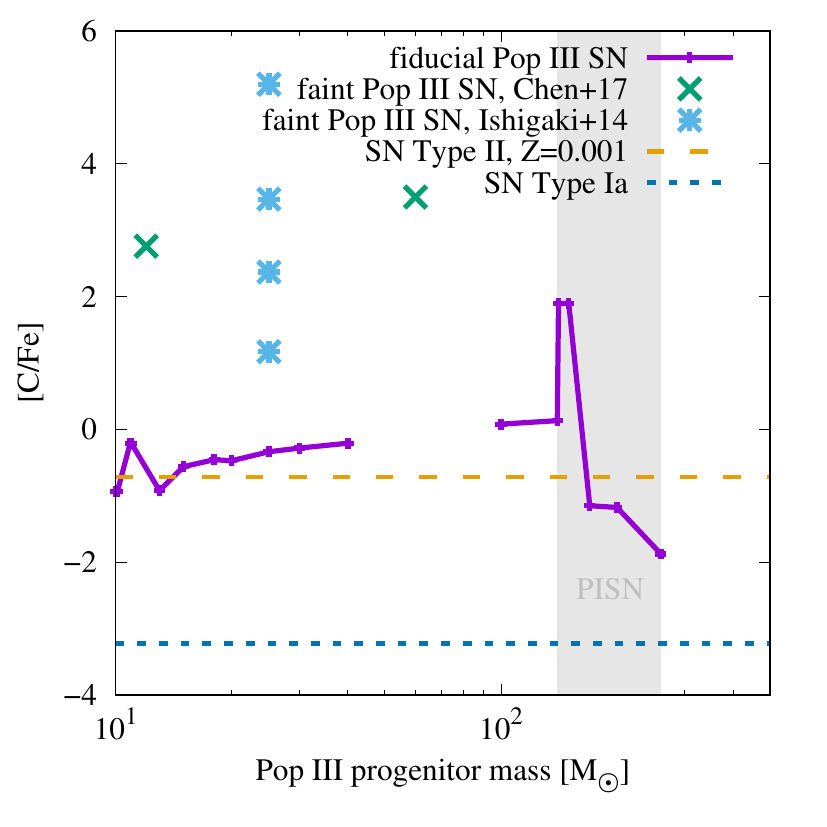}%{C-over-Fe}
\caption{Carbon-to-iron ratio, [C/Fe], as a function of the Pop~III SN progenitor mass \citep[solid,][]{nomoto13}. For comparison, we also show the yields of Type Ia \citep[short-dashed,][]{seitenzahl13} and Type~II SNe \citep[long-dashed,][]{nomoto13}. The yields for individual faint SNe are based on \citet{chen17} and \citet{ishigaki14}. PISNe with a progenitor mass of $\sim 150\Msun$ yield a very high [C/Fe] (because they eject relatively little iron), but PISNe with a progenitor mass of $\sim 250\Msun$ yield a very low, even significantly subsolar value of [C/Fe]. The explosion energies of Type~II SNe are assumed to be $10^{51}$\,erg. Faint SNe with lower explosion energies have generally higher [C/Fe] because more iron falls back onto the compact remnant. }
\label{fig:CFe}
\end{figure}

One important observed characteristic of extremely metal-poor stars is their frequently high carbon-to-iron ratio, which we aim to reproduce in our model by including faint SNe. We illustrate the [C/Fe] ratio as a function of Pop~III progenitor mass in Fig.~\ref{fig:CFe} for different types of SNe.
A faint SN refers to an explosion with a very small ejected $^{56}$Ni mass either due to a low explosion energy \citep{chen17} or large-scale mixing and fallback in aspherical explosions \citep{tominaga07}. To account for faint SNe, we include the corresponding yields by \cite{ishigaki14} in our model and discuss the calibration of the fraction of faint SNe in Section \ref{sec:calib}. These models are all for faint SNe with a progenitor mass of $25\Msun$, but can be taken as representative for faint SNe occurring in the mass range $10-40\Msun$.

\section{Results}

\subsection{Calibration}
\label{sec:calib}
We use the observed fraction of carbon enhanced metal-poor (CEMP) stars and the distribution of EMP halo stars to calibrate our model. However, our model is not intended to reproduce these functions over a broad metallicity range because we focus on second generation stars. In general, metal-poor stars can form after any number of previous generations of star formation, but each additional enrichment event results in higher stellar metallicities. Therefore, we focus on the stars with a metallicity of [Fe/H]$\leq -3$ for calibration purposes because we can assume that Pop~III stars dominate the enrichment of these EMP stars. The fraction of CEMP stars might be an inherent signature of the metal yields of Pop~III stars \citep{frebel07,cooke11,norris13,lee13,cooke14,placco14,bonifacio15,maeder15,caffau16,jeon17}, and thus less affected by any missing contribution from later generations.

\subsubsection{Fiducial model}
In this section, we present our fiducial parameters, motivate that they are physically reasonable, and that they meet additional observational constraints. Throughout the paper, we restrict our analysis to the MW and satellites within $R_\mathrm{vir}=300$\,kpc from the MW centre at $z=0$ (if not explicitly stated otherwise).%to avoid boundary effects due to missing radiative and chemical feedback from halos outside of our simulated volume.

The main model parameters and their fiducial values are summarized in Table~\ref{tab:fiducial}.
\begin{table}
 \centering
\begin{tabular}{ll}
Parameter & Value \\ 
\hline 
mass threshold for Pop~III & Eq. \ref{eq:mcrit} \\ 
mass threshold with LW feedback & Eq. \ref{eq:LW} \\ 
Pop~III SFE & $\eta_{\rm III}=0.001$ \\ 
Pop~II SFE & $\eta_{\rm II}=0.01$ \\ 
fraction of faint SNe & $f_\mathrm{faint}=40\%$ \\ 
metal fallback fraction & $f_\mathrm{fallback}=20\%$ \\ 
metal ejection fraction & $f_\mathrm{eject}=80\%$ \\ 
Pop~III SN wind velocity & $v=10$\,km/s \\ 
lower IMF limit & $M_\mathrm{min}=3\Msun$ \\ 
upper IMF limit & $M_\mathrm{max}=150\Msun$ \\ 
recovery time & $t_\mathrm{recov}=100\,$Myr \\ 
mean of dilution distribution & $\mu=10^{-1.5}$ \\ 
width of dilution distribution & $\sigma=0.75$\,dex \\ 
\end{tabular} 
  \caption{Parameter values in our fiducial model. This set of parameters best reproduces observations at [Fe/H]$\leq -3$ as we show below.}
   \label{tab:fiducial}
\end{table}
The Pop~III SFE is a crucial parameter for stellar archaeology since it defines the gas mass fraction that turns into stars and hence the average number of Pop~III SNe per minihalo. As well as calibrating it with stellar archaeology observations, we also enforce two additional constraints. We require that our choice of $\eta_\mathrm{III}$ leads to an optical depth for the Thomson scattering of CMB photons, $\tau=0.069$, consistent with the value measured by \citet{planck15}. See \citet{hartwig15} for a more detailed discussion on how the ionisation history of the Universe can be used to calibrate the Pop~III SFE in semi-analytical models. We also confirm that the mass in Pop~III stars per minihalo implied by our adopted SFE is consistent with the values found in detailed hydrodynamical simulations of Pop~III star formation. For example, for a minihalo with a total mass of $3 \times 10^{6} \: {\rm M_{\odot}}$ (the lowest mass minihalo capable of forming Pop~III stars at redshift $z \sim 20$), our fiducial Pop~III SFE predicts a total Pop~III stellar mass of around $500 \: {\rm M_{\odot}}$, in good agreement with the values of order $100-1000\Msun$ found in numerical simulations \citep{susa14,hirano14}. These numerical results can be seen as a lower limit because most simulations focus on the first high redshift peaks but we also expect metal-free star formation at $z<10$ in more massive halos. The fractions of ejected metals and metals that fall back onto the halo after an SN are consistent with the results of \citet{ritter12}.

\begin{figure}
\centering
\includegraphics[width=0.47\textwidth]{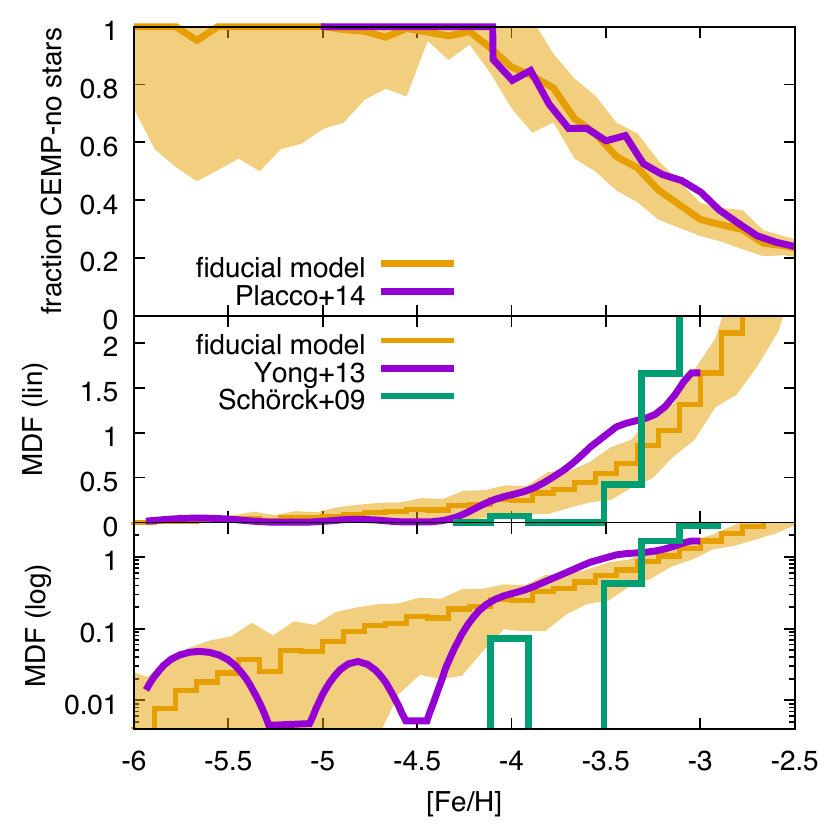}%{MDF_fiducial}
\caption{Top: Fraction of CEMP-no stars ([C/Fe]$>0.7$) as a function of [Fe/H] predicted by our model (orange) with the observed distribution \citep[purple,][]{placco14} shown for comparison. Below: Predicted (orange) and observed (green: \citealt{schork09}; purple: \citealt{yong13}) metallicity distribution functions, normalised to the number of stars below [Fe/H]$\leq -3$ in linear (middle) and logarithmic (bottom) scaling. The shaded regions indicate the scatter over 30 different merger tree realizations. Our model agrees with the observed distributions from \citet{placco14} and \citet{yong13} within the statistical uncertainty.}
\label{fig:calibration}
\end{figure}

We show in Fig.~\ref{fig:calibration} that we can reproduce the metallicity distribution function (MDF) and the fraction of CEMP-no stars as a function of metallicity with this set of parameters. CEMP-no stars are a subclass of CEMP stars with [Ba/Fe]$\leq 0.0$, i.e.\ with no enhancement in neutron capture elements. %with no significant contribution from neutron capture elements. APJ edit
We limit this comparison to stars with [Fe/H]$\leq -3$ because above this value we expect contributions from later generations of stars to become important that we do not model self-consistently. In the low metallicity range, we can successfully reproduce both observed distributions with our model. We used the fraction of CEMP-no stars with [C/Fe]$>0.7$ from \citet{placco14} and the MDF from \citet{yong13} for comparison, since the latter is more recent and complete than the MDF provided by \citet{schork09}. Since we average 30 MW-like merger trees we do not reproduce the sparse sampling at [Fe/H]$\leq -4.5$, but discuss this effect separately below.

Another important and poorly constrained parameter is the fraction of faint SNe, which is assumed to have a direct influence on the fraction of CEMP stars due to the high [C/Fe] yields of this type of SN. We find a best matching value of $f_\mathrm{faint}=40\%$. Slightly higher values \citep{ji15,deB17} are also compatible within our error margins. The fraction of CEMP stars is mainly controlled by the adopted model for mixing with the ISM and $f_\mathrm{faint}$.

\subsubsection{Exploring input parameters}
\begin{figure}
\centering
\includegraphics[width=0.47\textwidth]{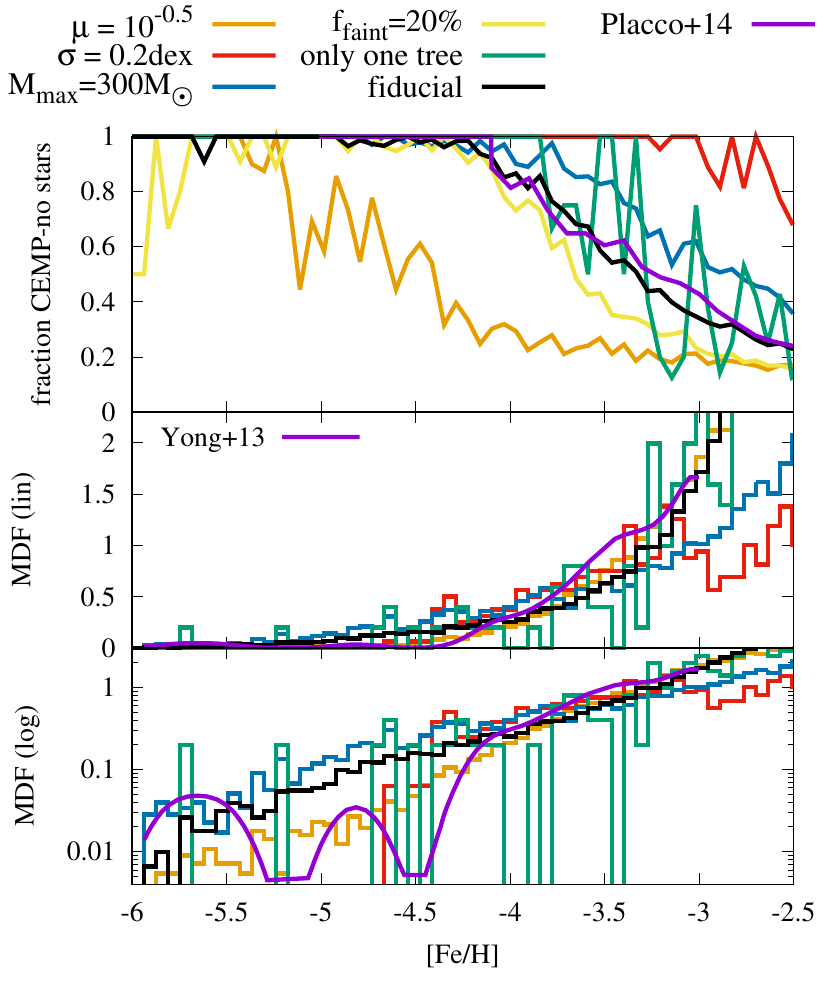}%{MDF_para}
\caption{Same as Fig.~\ref{fig:calibration}, but showing the effect of varying the model parameters specified in the legend. The results of our fiducial model are shown in black. We also show the results from a realization based on only one tree (green), which highlights the expected stochasticity of the distribution at low [Fe/H]. We note that the yellow line in the middle and lower panels ($f_\mathrm{faint}=20\%$) is identical with that in the fiducial model.}
\label{fig:para}
\end{figure}

We now compare how different parameters affect the results and demonstrate quantitatively that our fiducial set of parameters best reproduces the MDF and the fraction of CEMP-no stars (Fig.~\ref{fig:para}).
If we assume $\mu=10^{-0.5}$, i.e. that metals ejected by SNe mix with almost all available hydrogen in a halo, we predict too few CEMP-no stars. If we assume that the distribution of dilution factors is too narrow ($\sigma=0.2$\,dex), we predict too many CEMP-no stars with [Fe/H]$\approx -3$. The green line in this plot also demonstrates that our model for a single MW-like merger tree correctly reproduces the sparsely sampled region at [Fe/H]$\leq -4.5$.

To quantify the quality of our calibration and to compare the relative influence of the model parameters in Table \ref{tab:fiducial}, we apply the Kolmogorov-Smirnov (KS) test and calculate the maximum difference between the cumulative distribution functions of the two observed distributions and our model:
\begin{equation}
D=\max _{x \leq -3} |F_\mathrm{obs}(x)-F_\mathrm{model}(x)|,
\label{eq:KS}
\end{equation}
where $F(x)$ is the cumulative distribution function and $x=$[Fe/H]. The resulting values for various models are summarized in Table \ref{tab:KS}.
\begin{table}
 \centering
\begin{tabular}{llll}
Parameter & $D_\mathrm{MDF}$ & $D_\mathrm{CEMP}$ & $\Sigma$ \\ 
\hline 
fiducial & $0.08$ &$0.07$ & $0.15$ \\
$M_\mathrm{min}=10\Msun$ & $0.09$ & $0.11$ & $0.20$ \\ 
$M_\mathrm{max}=120\Msun$ & $0.13$ & $0.11$ & $0.24$ \\
$M_\mathrm{max}=300\Msun$ & $0.24$ & $0.07$ & $0.31$ \\
$\eta_{\rm III}=0.0005$ & $0.14$ & $0.05$ & $0.19$ \\
$\eta_{\rm III}=0.002$ & $0.12$ & $0.10$ & $0.22$ \\
$\eta_{\rm II}=0.02$ & $0.12$ & $0.06$ & $0.18$ \\
$t_\mathrm{recov}=10$\,Myr & $0.13$ & $0.05$ & $0.18$ \\
$f_\mathrm{faint}=0.2$ & $0.08$ & $0.12$ & $0.20$ \\
$f_\mathrm{faint}=0.6$ & $0.08$ & $0.08$ & $0.16$ \\
IMF slope: $-1$ & $0.14$ & $0.18$ & $0.32$ \\
$f_\mathrm{eject}=0.5$ & $0.14$ & $0.07$ & $0.21$ \\
$\mu=10^{-0.5}$ & $0.07$ & $0.15$ & $0.22$ \\
$\mu=10^{-2.0}$ & $0.09$ & $0.07$ & $0.16$ \\
$\sigma=0.2$\,dex & $0.16$ & $0.16$ & $0.32$ \\
\end{tabular} 
  \caption{Parameter study and KS-test values (Eq. \ref{eq:KS}). Our fiducial model yields the smallest maximum differences between the cumulative distributions of the observations and our model. However, the only model that can be rejected based on this two-sample KS test at the 95\% level is the one with $M_\mathrm{max}=300\Msun$ (see text).}
   \label{tab:KS}
\end{table}
Our fiducial set of parameters minimizes the sum of $D_\mathrm{MDF}$ and $D_\mathrm{CEMP}$. To reject the null-hypothesis that our model reproduces the observations at 95\% significance level, we determine the corresponding critical distance to be $D_\mathrm{crit,MDF}=0.23$ for the MDF and $D_\mathrm{crit,CEMP}=0.29$ for the fraction of CEMP-no stars. The only parameter choice that can be excluded based on this analysis is $M_\mathrm{max}\geq 300\Msun$ as an upper limit for the Pop~III IMF. Since we do not fully explore our 11D parameter space, we can only conclude that our fiducial parameters represent a local optimum, while other parameter combinations may yield a similar or even better fit to the observations. Unfortunately, this also illustrates the weak predictive power of this approach and we do not claim to constrain any of the parameters by fitting a model with 11 free parameters to two observables. A full parameter space exploration could be performed by means of, e.g., Gaussian processes model emulators \citep[e.g.][]{2010MNRAS.407.2017B,2012ApJ...760..112G, 2014ApJ...787...20G}.
%This technique is commonly used to generate model emulators that allow one to statistically estimate a desired set of model outputs at any location of a p-dimensional input parameter space \citep[e.g.][]{2010MNRAS.407.2017B,2012ApJ...760..112G, 2014ApJ...787...20G}. As a result, one can explore  the full input parameter space orders of magnitude faster than could be done otherwise.
Nonetheless, our set of initial parameters agrees with other studies and reproduces the main observations provided by stellar archaeology. Moreover, we will show later that our main conclusions can also be derived independently of the specific cosmological model adopted.

%We find a logarithmically flat IMF to provide the best fit to observations and that a slope of $\mathrm{d}M/\mathrm{d},\\mathrm{log}M = -1$ fails, especially at reproducing the observed fraction of CEMP-no stars. However, \citet{deB17} explicitly show with a similar semi-analytical merger tree-based approach that a logarithmically flat Pop~III IMF is disvavored because it fails to reproduce the low-metallicity tail of the MDF. excludes a logarithmically flat Pop~III IMF. Our main conclusions independent of the details of the cosmological model.

\begin{figure}
\centering
\includegraphics[width=0.47\textwidth]{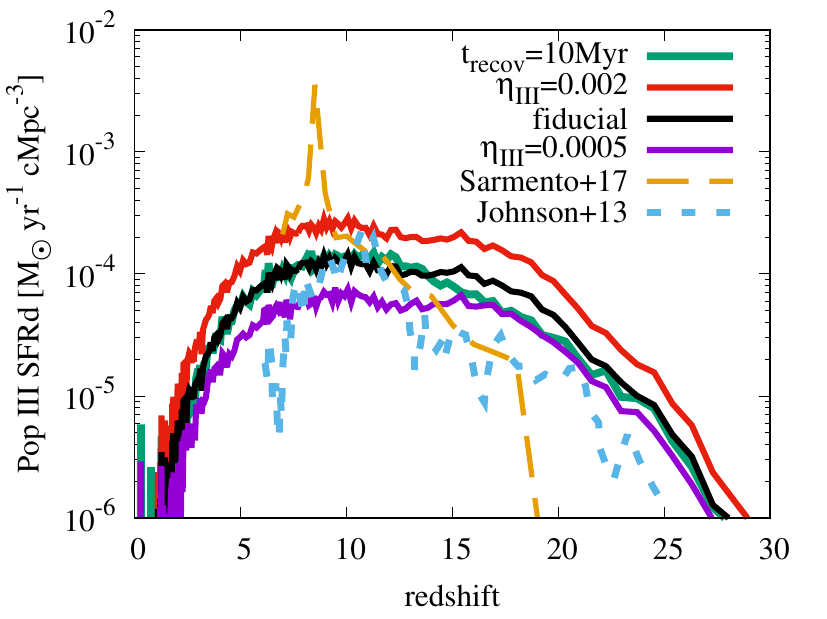}%{z_SFR}
\caption{Comparison of the SFRd for Pop~III stars as a function of redshift of our model (solid) to the rates by \citet{johnson13} and \citet{sarmento17} (dashed). The SFRd scales roughly with the SFE, and in our fiducial model, we find a peak value of $\sim 10^{-4}\Msun\,\mathrm{yr}^{-1}\,\mathrm{cMpc}^{-3}$ (co-moving Mpc) around $z\approx 10$. A shorter recovery time leads to a more efficient suppression of Pop~III SF at $z \gtrsim 15$ because Pop~II stars can form earlier. The SFRd of our model is averaged over 30 MW-like trees.}
\label{fig:zSFR}
\end{figure}

We also show the parameter dependence of the Pop~III star formation rate density (SFRd) in Fig.~\ref{fig:zSFR}.
It is calculated within the co-moving volume of the MW and therefore represents a cosmic overdensity. Our star formation rates are consistent with those in \citet{johnson13}, with the upper limit advocated by \citet{visbal15}, and with the Thomson scattering optical depth measured by \citet{planck15}. Our results differ from \citet{sarmento17} because they allow Pop~III star formation in slightly enriched halos up to a metallicity of $Z_\mathrm{crit}=10^{-5}Z_\odot$, which permits more Pop~III star-forming halos at $z<10$.

\subsection{Internal vs. external enrichment}
\begin{figure}
\centering
\includegraphics[width=0.47\textwidth]{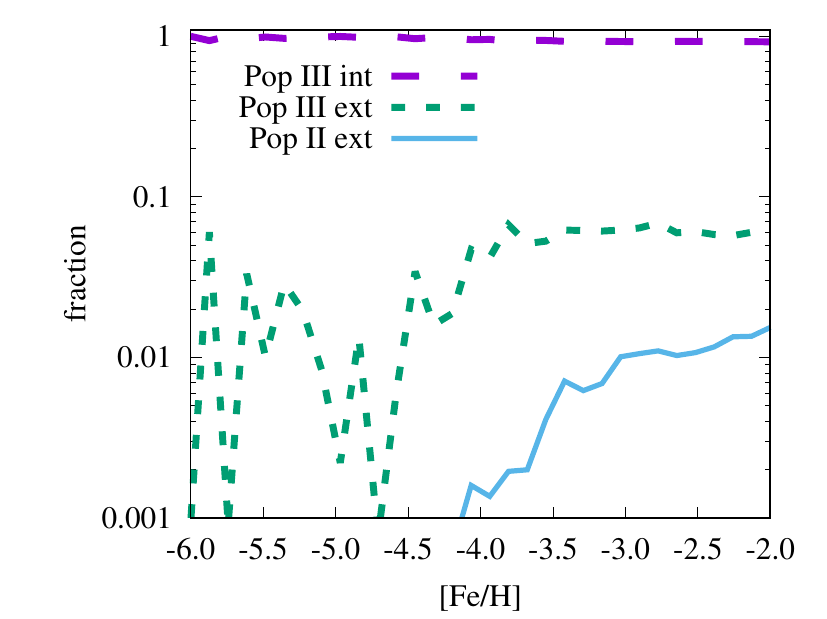}%{MDF_rel_fiducial}
\caption{Relative contribution to the metal enrichment of second generation stars via different enrichment channels (metal mass weighted). The three contributions sum up to $100\%$. Internal enrichment by Pop~III stars dominates at all metallicities and external enrichment by Pop~III stars accounts for $\sim 10\%$ of the enriching metals above [Fe/H]$=-4$. External enrichment by Pop~II stars is always sub-dominant ($\lesssim 1\%$) for the overall metal budget of second generation stars.}
\label{fig:MDFrel}
\end{figure}

The difference between internal and external enrichment is important because the timescales of the subsequent collapse and the overall enriching mass depend on the nature of the enrichment. As internal enrichment, we label the inevitable chemical enrichment of a halo after star formation. External enrichment occurs when the radius of a metal-enriched bubble is larger than the separation between the centres of two halos (see Sec. \ref{sec:yields}), typically of the order $0.1-10$\,kpc. Both of these enriching events are passed through the merger tree so that a halo at $z=0$ could have experienced several internal and external enrichment events during its assembly history. We investigate the relative contributions of internal vs. external enrichment in Fig.~\ref{fig:MDFrel}.
Internal enrichment is dominant compared to external enrichment prior to the formation of second generation stars, as has also been shown by \citet{griffen17}, \citet{visbal17}, and \citet{jeon17}. If halos are close enough for external enrichment, ionizing feedback is usually also strong enough to suppress star formation, thereby preventing the formation of externally enriched second generation stars. For this reason, varying the recovery time makes little difference to the external enrichment fractions. The metal contributions in Fig.~\ref{fig:MDFrel} are averaged and there are individual halos that are only enriched externally by Pop~III or Pop~II stars, although their occurrence in number is small. We find that the outcome of second generation star formation does not strongly depend on environmental effects, such as the clustering of halos. We also confirm in our semi-analytical model that the radial distribution of halos hosting second generation stars follows the radial distributions of all halos in the local volume at $z=0$.

% (related to distance figure of mass binned halos of Griffen+17)?
%[compare to 3D simulations; are there cosmologically representative simulations that have studied this? Because \citet{wise12,ritter12,smith15} only focus on single halos and not on a statistical average.]

\begin{figure}
\centering
\includegraphics[width=0.47\textwidth]{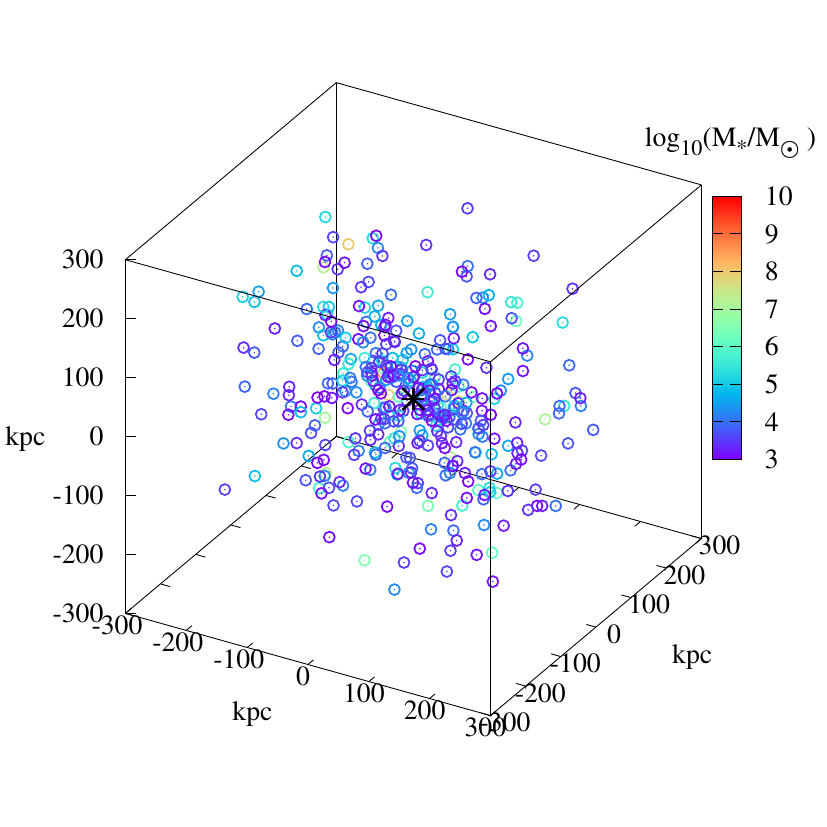}%{r_M}
\caption{Projection of all star-hosting halos at $z=0$ within 300\,kpc of the MW main halo for one merger tree realization. The main halo is indicated by the black asterisk and the satellites are colour-coded by their stellar mass.}
%\caption{Radial distribution of all star-hosting halos at $z=0$ within $300$\,kpc of the MW main halo for one merger tree realization. The stellar masses are assigned via abundance matching, based on the DM peak mass. The right y-axis and the purple line indicate the cumulative fraction of MW satellites.}
\label{fig:3D}
\end{figure}

In Fig.~\ref{fig:3D}, we see the 3D distribution of halos in the local group at $z=0$ for one exemplary MW-like merger tree.
We find $\sim 400$ satellites with stellar masses above $1000\Msun$. The observed number of MW satellites is around 50 \citep{drlica15}, which seems to be in contradiction with our model and other DM simulations \citep[the ``missing satellite problem'', see][]{kauffmann93,klypin99,moore99}. However, this discrepancy can be solved by correcting for the completeness bias of the surveys \citep{kim17}. We assign stellar masses at $z=0$ via abundance matching based on the peak mass of each satellite during its assembly history \citep{gk14}. Note that stellar masses below $\sim 5 \times 10^{5}\Msun$ should be considered as an extrapolation due to the incompleteness of their observations for low mass satellites. Moreover, the scatter in the relation between stellar and halo mass becomes more important at lower masses \citep{gk17}. Hydrodynamic simulations indicate that extrapolations to low masses are reasonable \citep{munshi17,jeon17}, but our stellar masses at $z=0$ should be considered as rough estimate for lower-mass satellites.

\begin{figure}
\centering
\includegraphics[width=0.47\textwidth]{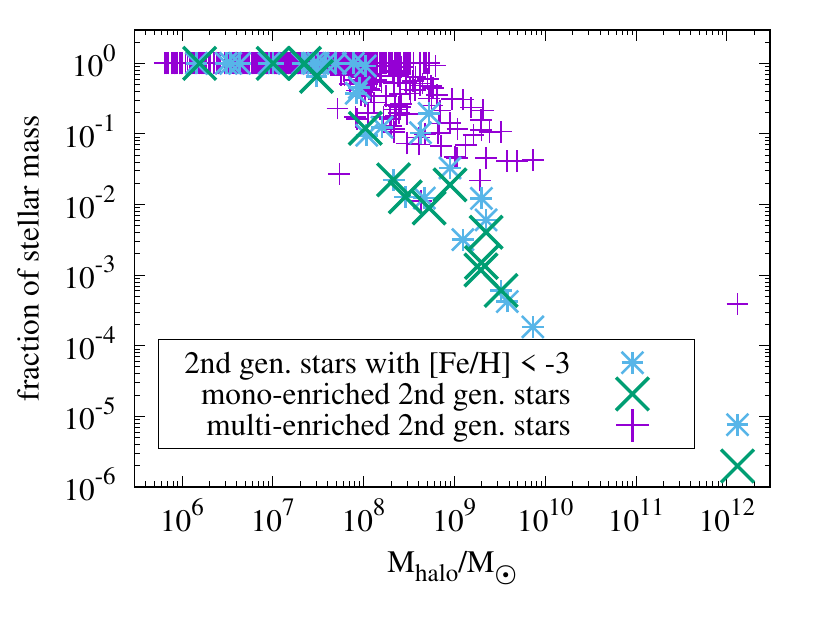}%{MM}
\caption{Fraction of all (purple), metal-poor ($[{\rm Fe}/{\rm H}] < -3$, blue), and mono-enriched (green) second generation stars relative to the total stellar mass at $z=0$. Second generation stars end up in satellites of all masses, but their fraction is much higher in low mass halos.}
%Some satellites with $M_{\rm h} \lesssim 10^8\Msun$ consist mostly of second generation stars.}
\label{fig:MM}
\end{figure}

For a direct comparison of the fractions of second generation stars we assume for the mass of the stellar population of the second generation an instantaneous starburst which converts $1\%$ of the gas mass into stars.
%This simplistic model for metal-enriched star formation reproduces the stellar mass of the MW of $(3-7)\times 10^{10}\Msun$ at $z=0$ \citep{flynn06,McM11,bovy13,licquia15}.
The resulting fractions as a function of the stellar mass can be seen in Fig.~\ref{fig:MM}.
During the assembly of the MW and its satellites, halos that host second generation stars merge into larger systems and at $z=0$ second generation stars can be found in satellites of all masses. However, the relative contribution of second generation stars to the total stellar population depends on the host mass, with less massive halos being more likely to host a higher fraction of second generation stars. The MW at $z=0$ consists of e.g.\ $\lesssim 0.1\%$ second generation stars, but only $\sim 10^{-5}$ of all MW stars are metal-poor ([Fe/H]$<-3$) and $\sim 10^{-6}$ are mono-enriched second generation stars. Our analysis shows that the stellar population in satellites with $M_{\rm h} \lesssim 10^8\Msun$ originates dominantly from the second generation of star formation. Although our model predicts a fraction of close to $100\%$ in this mass range, the actual fraction may be lower due to the scatter in the halo to stellar mass relation, which we do not take into account.

These results are in agreement with previous models that show that ultra-faint dwarf galaxies host ancient stellar populations and probe early cosmic star formation \citep{bullock00,salvadori09,gao10,starkenburg13,weisz14,ji15b,jeon17,griffen17,starkenburg17b}. This is because ultra-faint dwarf galaxies with $M_h < 2 \times 10^9\Msun$ formed $\gtrsim 90\%$ of their stellar mass prior to reionization \citep{jeon17} and have an average iron abundance of [Fe/H]$<-2$ \citep{kirby08}.
% APJ: Note that Brown et al 2014 find >80% at z > 6, worth including this.

\subsection{Number of Pop~III SNe per halo}
\label{sec:NSN}
\begin{figure}
\centering
\includegraphics[width=0.47\textwidth]{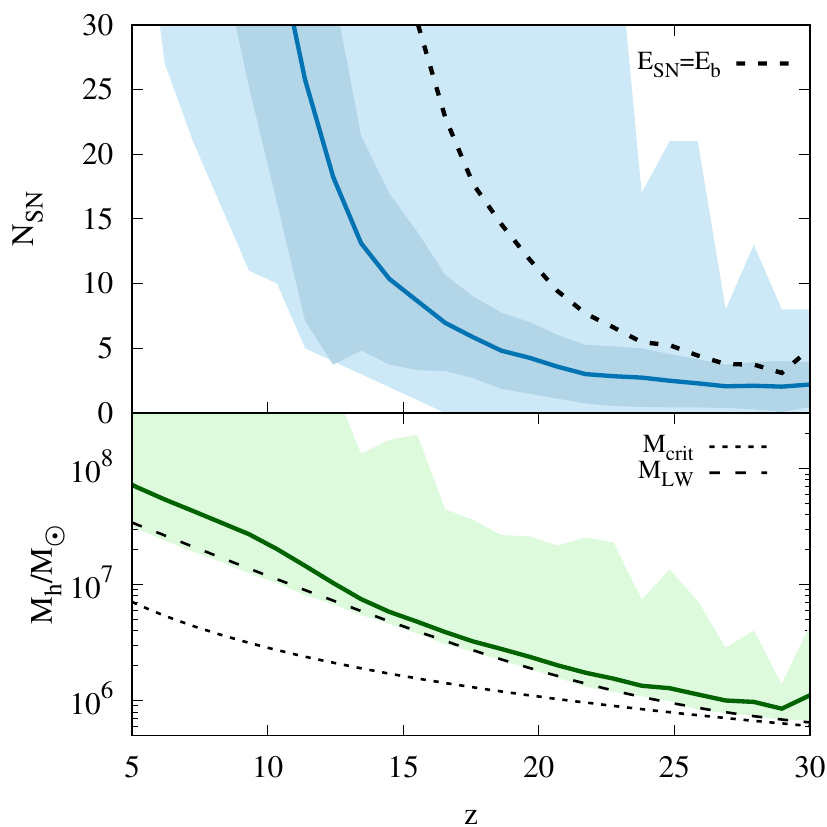}%{z_NSN_Mcrit}
\caption{Top: Number of Pop~III SNe per minihalo as a function of redshift. Bottom: Halo masses at the moment of Pop~III star formation. The solid line indicates the mean, the dark contours the $1\sigma$ standard deviation, and the light contours the minimum and maximum values in this redshift bin. The increase of the number of SNe with decreasing redshift is related to the simultaneous increase of the stellar mass that is available per Pop~III star-forming halo. In some rare cases at $z>15$ there are minihalos with only one SN, but generally we expect between 5 and 20 SNe per minihalo. The dotted and dashed lines in the bottom panel illustrate the critical masses for Pop~III star formation. The dotted line in the top panel indicates the number of SNe required to expel all of the gas from the halo. Halos with more than this number of SNe may be completely disrupted by Pop~III SNe and hence may not form second generation stars.}
\label{fig:zNSN}
\end{figure}

The chemical signature of second generation stars can be used to deduce the masses of their Pop~III progenitors. For this purpose, we are especially interested in those cases where the ISM was enriched by exactly one previous Pop~III SN. However, in most minihalos we form Pop~III stars in small multiples \citep{turk09,stacy10,clark11,greif11b,smith11,susa14,hirano17b} and in Fig.~\ref{fig:zNSN}, we show the average number of SNe per minihalo.
It is an increasing function with decreasing redshift due to the increasing threshold mass for Pop~III star formation. At $z\gtrsim 15$ we expect fewer than 10 SNe per halo and in individual cases there are halos with just one Pop~III SN. These are the cradles for mono-enriched second generation stars.

Minihalos at high redshift have shallow potential wells and SNe could unbind all the gas in the halo and hence prevent subsequent star formation. To derive the critical number of SNe required to do this, we assume that an SN has on average an energy of $10^{51}$\,erg and that the halo has a gravitational binding energy \citep{loeb10} of
\begin{equation}
E_\mathrm{b}=2.9 \times 10^{53} \left( \frac{M_h}{10^8 \Msun} \right)^{5/3} \left( \frac{1+z}{10} \right) \,\mathrm{erg}.
\end{equation}
Not all of the injected SN energy will effectively couple to the gas and contribute to its ejection, as some will instead be radiated away. Also the low-density H{\sc ii} region, which surrounds the first stars at the moment of their SN explosions, and the anisotropy of the ISM, which provides channels of least resistance for the energy to escape, reduce the efficiency of SNe in ejecting gas from the galactic potential well. To account for this effect, we assume that only 10\% of the SN energy couples efficiently to the gas \citep{kitayama05,whalen08}. This yields the number of SNe per halo that is required to unbind all gas as
\begin{equation}
N_\mathrm{SN}=62 \left( \frac{M_h}{10^7 \Msun} \right)^{5/3} \left( \frac{1+z}{10} \right).
\end{equation}
The black dashed line in the upper panel of Fig.~\ref{fig:zNSN} indicates that this critical value is above the average number of SNe per halo. Nevertheless, some halos at every redshift have values of  $N_\mathrm{SN}$ above this critical value, and may therefore form fewer multi-enriched second generation stars than our model assumes, because of the disruption of these halos by SN feedback. We note, however, that this is a simplistic order of magnitude estimate and more realistic models show that gas fallback is also possible after several or more energetic SN explosions in a minihalo \citep{kitayama05,greif10,ritter12,chiaki18}. Therefore, we do not include this destructive effect of multiple SNe self-consistently in our model, but highlight possible implications in the discussion section.

\begin{figure}
\centering
\includegraphics[width=0.47\textwidth]{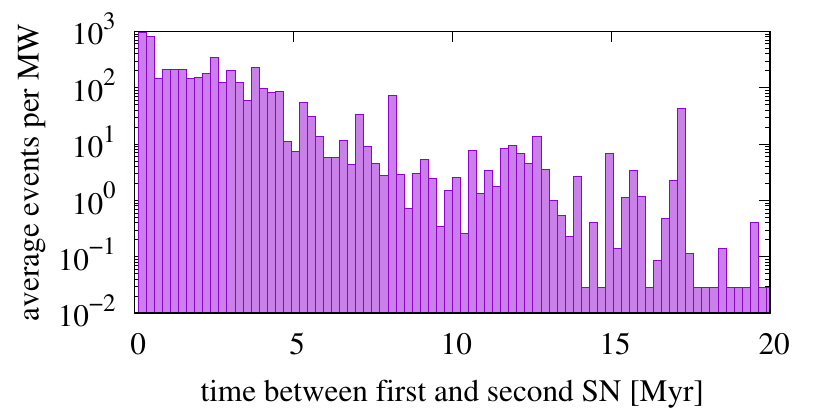}%{histo_dtSN}
\caption{Histogram of the times between the explosion of the first and the second SN in minihalos per MW-like merger tree. Due to the very short lifetimes of massive stars, the second SN explodes generally within less than 10\,Myr after the first one (mind the logarithmic y-axis). This is shorter than the typical recovery time for second generation star formation ($\sim 100$\,Myr). The dominance of short times between SNe illustrates that there is generally not enough time between two SN explosions to form second generation stars. Instead, they form after most of the Pop~III stars in the minihalo have exploded as SNe.}
\label{fig:dtSN}
\end{figure}

It is also interesting to examine whether the time between two SNe is long enough for the gas to recollapse and form mono-enriched second generation stars before the second SN explodes. In Fig.~\ref{fig:dtSN} we show a histogram of the times between the explosion of the first and the second SN in minihalos.
The average time between two SNe is much shorter than our assumed recovery time for second generation star formation. Consequently, the presence of multiple SNe in one minihalo indicates that most stars that form at the onset of Pop~II star formation carry the imprint of several Pop~III SNe.
%We emphasis that our model of Pop~III star formation is purely statistical and we do not follow the protostellar collapse and the fragmentation of the disk.

\begin{figure}
\centering
\includegraphics[width=0.47\textwidth]{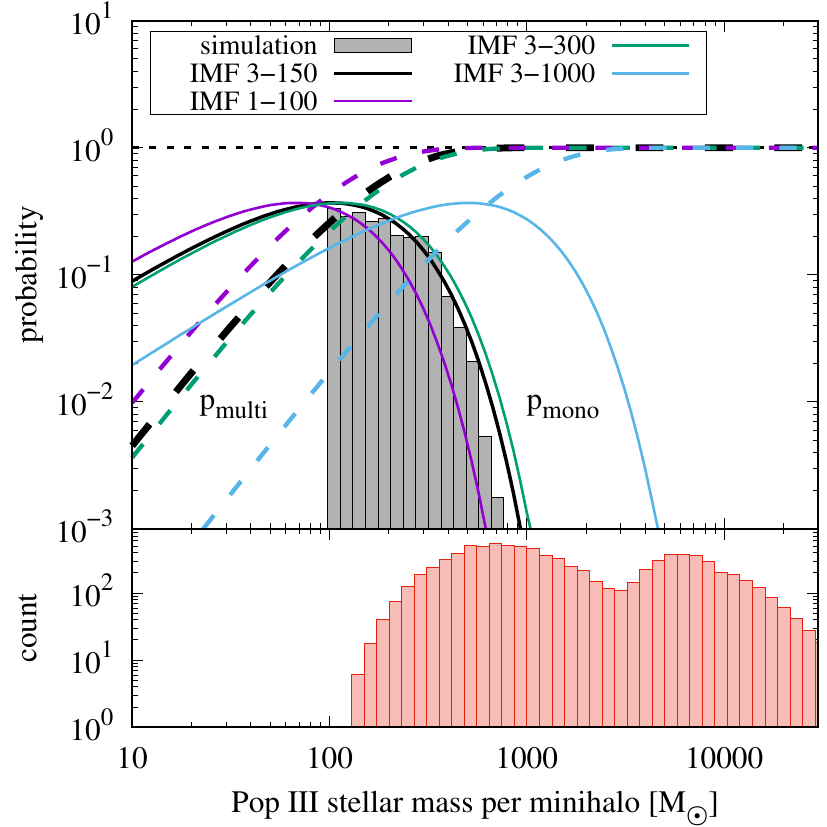}%{Poisson_SNe}
\caption{Top: Probability to have exactly one SN (solid) or more than one SN (dashed) per minihalo as a function of the stellar mass for different IMF ranges. The black lines correspond to the analytical prediction of our fiducial model and should be compared to the grey histogram, which is the average over all 30 merger trees. Bottom: histogram of the stellar masses of Pop~III star-forming halos in one MW-like realization. Most Pop~III stars form in minihalos with $M_* \lesssim 1000\Msun$ but some form in atomic cooling halos with stellar masses up to $M_*\gtrsim 10^4\Msun$. In these mass ranges the probability to have exactly one SN in a randomly selected minihalo is $< 1\%$.}
\label{fig:NSN}
\end{figure}

We derive the probability that exactly one SN explodes in a minihalo, based on Poisson statistics. For a given Pop~III IMF we calculate how much stellar mass we need on average to form one SN.
%For a logarithmically flat IMF in the mass ranges 1-100$\Msun$, 3-300$\Msun$, and 10-1000$\Msun$ we expect on average one SN per $68\Msun$, $114\Msun$, and $505\Msun$ of stellar mass, respectively.
The mean number of SNe in a halo with stellar mass $M_*$ is then given by
\begin{equation}
\lambda = \frac{M_*}{M_\mathrm{1SN}},
\end{equation}
where $M_\mathrm{1SN}$ is the stellar mass to expect on average one SN. By applying Poisson statistics, we calculate the probability to have $k$ SNe going off in one minihalo:
\begin{equation}
p(k) = \frac{\lambda ^k}{k!} \mathrm{e}^{- \lambda}.
\label{eq:pmono}
\end{equation}
The probability to have one SN per halo is given by $p(1)$ and the probability to have more than one SN per halo is given by $1-p(0)-p(1)$. These probabilities can be seen as a function of the stellar mass in Fig.~\ref{fig:NSN}.
This analytical derivation is valid as long as the total stellar mass is higher than the upper IMF limit because otherwise the entire IMF cannot be sampled. As we can see in the bottom panel, this criterion is almost always fulfilled in our fiducial model because we form at least $\sim 100\Msun$ of Pop~III stars per halo (Eq. \ref{eq:Mstar}). Consequently, the probability to have only one SNe per minihalo is very low, of the order 1\%. Instead, we expect second generation stars to form from gas that has been previously enriched by several SNe. This analytical estimate is very powerful and flexible because it predicts the probability of having more than one SN per minihalo for any possible IMF or stellar mass. The chances to create mono-enriched second generation stars are highest in the smallest minihalos because the available gas mass to form stars is lower and hence it is more likely for these halos to host  only one Pop~III star that explodes as an SN.

\section{Chemical Signature of Second Generation Stars}
We aim to find the optimal diagnostic and selection criteria for EMP stars that are promising mono-enriched candidates given that only relatively few elements are observable in EMP stars with reasonable effort. We thus need to quantify the likelihood for star-forming gas to have experienced only one prior enrichment event. We first use our semi-analytical model to find which abundances are best suited for this purpose. Then, we present a novel diagnostic that is independent of any model for primordial star formation and only depends on the assumed SN yields.

\subsection{Signature based on our cosmological model}
\begin{figure}
%\centering
\includegraphics[width=0.45\textwidth]{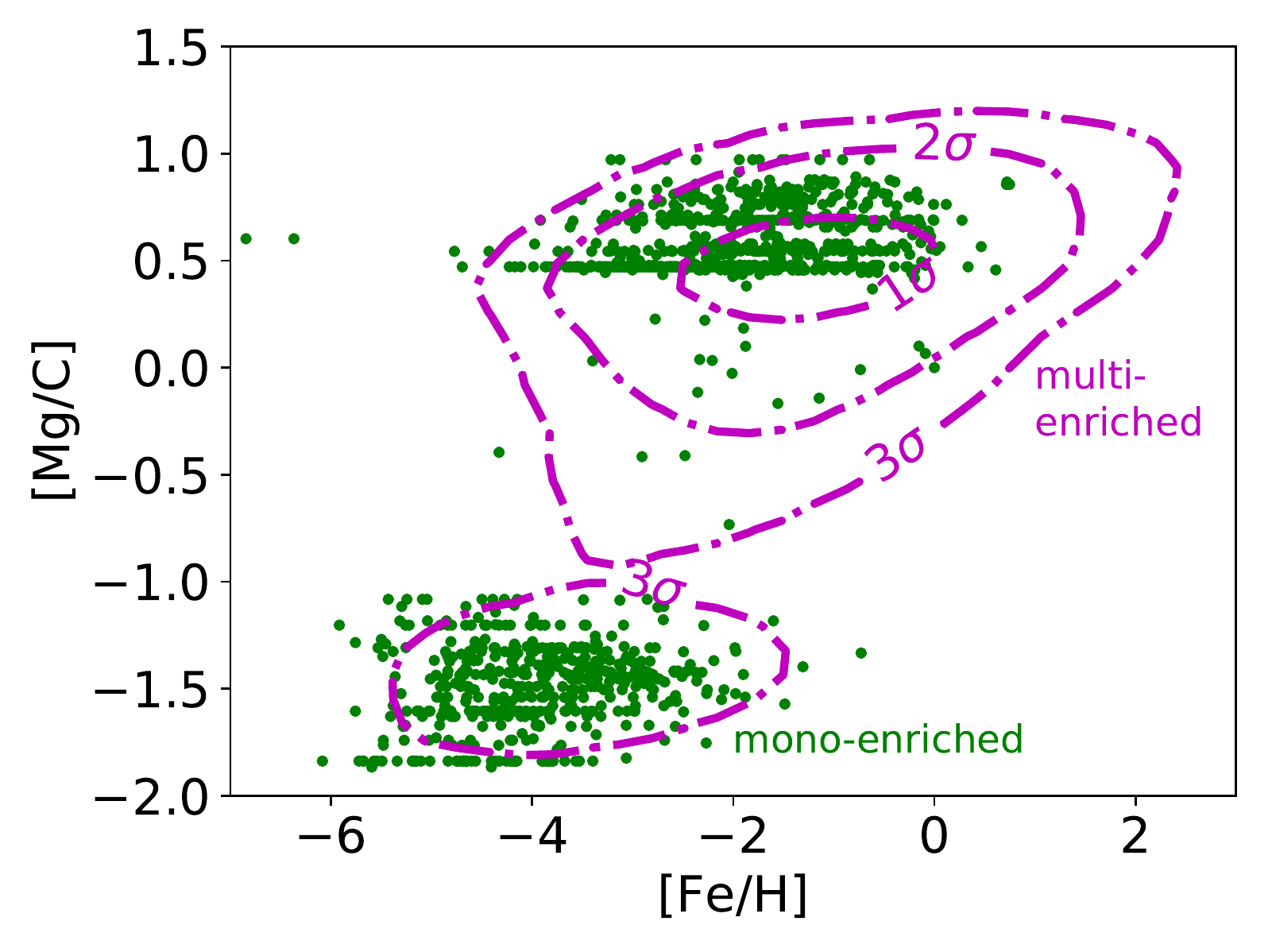} %{fiducial_Mg-over-C.eps}
\includegraphics[width=0.45\textwidth]{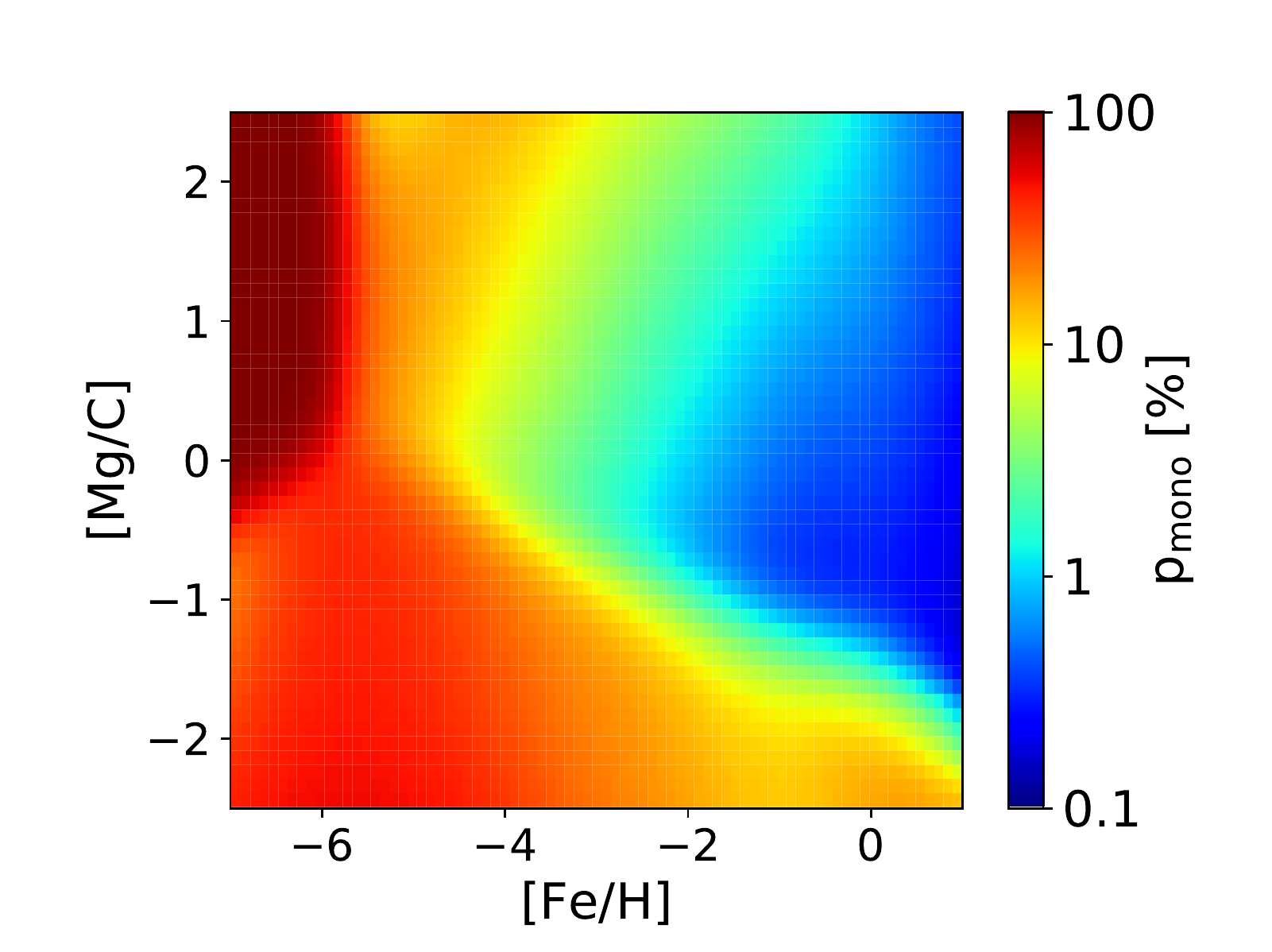} %smooth_probability_1SN.py
%\centering
\caption{Top: Mono-enriched second generation stars populate specific regions in this plot (green), compared to the distribution of multi-enriched second generation stars, illustrated by the purple probability contours. Mono-enriched stars can be found at all metallicities up to almost solar, although most have $[{\rm Fe}/{\rm H}] < -2$, and so the metallicity alone is not a reliable diagnostic for whether the star is mono-enriched or multi-enriched. [Mg/C] further helps to quantify the likelihood of the gas being enriched only once.
Bottom: Probability of mono-enrichment, $p=N_\mathrm{mono}/(N_\mathrm{multi}+N_\mathrm{mono})$, for the same elemental ratios as in the top panel.
There are regions of the parameter space in our model with a probability of almost 100\% for finding second generation stars that formed from gas that was enriched by only one previous SN. However, this probability does not reflect how many stars in total are expected in these regions, as we can see by comparing the two panels.}
\label{fig:pMgC}
\end{figure}

In Fig.~\ref{fig:pMgC}, we display as an example the distribution and probability of finding mono-enriched second generations stars, calculated for the [Mg/C] ratio. 
The mono-enriched second generation stars populate specific regimes, different from those of multi-enriched second generation stars. In general, the probability of mono-enrichment is a decreasing function of metallicity and we find even individual mono-enriched second generation stars with solar metallicities in our model. The abundance ratio [Mg/C] adds an additional constraint with the lowest probability for mono-enrichment around [Mg/C] $\sim 0$ and higher probabilities for higher and lower values of [Mg/C]. Such probability maps can be created for all abundance ratios and in higher dimensions. We limit the discussion to this two-dimensional representation to illustrate the concept since [Mg/C] can be observed with little effort and already provides a solid additional constraint.
%In our model, we then identify parameter ranges that lead to high probabilities of finding mono-enriched second generation stars, namely [Mg/C]$<-1$ and [Mg/C]$>0.5$.
%The region where it is very unlikely to find mono-enriched stars is [Mg/C] $\sim 0$.  
%We note that this example does not take into account uncertainties in the SN yields, or incomplete mixing of the elements within the ISM. These effects would effectively smooth the probability map, and eliminate any sharp boundaries between high and low probability regions signifying mono-enrichment.

We also take into account the theoretical uncertainty in the values of the SN yields and the typical observational uncertainties for derived stellar abundances. \citet{ishigaki18} compile the observational errors from recent high-resolution spectroscopic studies \citep{yong13,cohen13,roederer14} for the typically observed yields in EMP stars, which are in the range $0.1-0.5$\,dex, depending on the element and spectral resolution. \citet{nomoto13} compare the predicted metal yields from different groups for Pop~III SNe \citep{tominaga07,hw10,limongi12} and find a scatter between independent models of on average $0.3$\,dex for the elements carbon to zinc. We have additionally compared the theoretical yields from \citet{ishigaki18} to the predictions from \citet{hw10} and find a discrepancy for some elements of $>1$\,dex. Although the combined observational and theoretical uncertainty should be evaluated individually for every element, we assume for simplicity $0.5$\,dex, which is a reasonable average of the various sources of uncertainty. We consequently smooth the abundance-dependent distributions with a Gaussian convolution kernel with the width $\sigma = 0.5$\,dex to express that we cannot make exact predictions on finer scales.

We do not account for observational or theoretical uncertainties in the top panel of Fig.~\ref{fig:pMgC}. This is why the probability map in the lower panel extends to regions that are not sampled in the top panel. The two events at [Fe/H]$<-6$ and [Mg/C]$\sim 0.6$ correspond to a very small hydrogen dilution mass and a star in this region would have a probability close to 100\% to be mono-enriched (see lower panel). However, there are no observed stars in this abundance regime yet \citep{saga,abohalima17}.

%\subsection{Signature based on the divergence of the chemical displacement}
\subsection{The divergence of the chemical displacement}
In this section, we propose a new, alternative method to identify mono-enriched EMP stars based on their chemical composition. This method is independent of the star formation model, computationally efficient, and the qualitative results are insensitive to assumptions about the IMF or the fraction of faint SNe. We first introduce the underlying analytical arguments of this new diagnostic, compare it to the results from our cosmological model, and finally apply it to observed EMP stars.

\subsubsection{Motivation and Definition}
\begin{figure}
\centering
\includegraphics[width=0.47\textwidth]{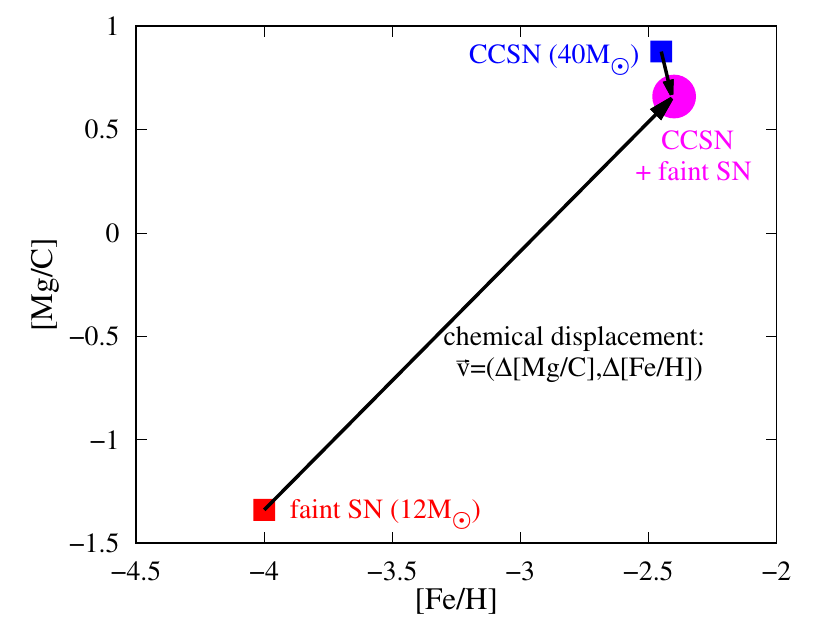}%{Div_illustration}%{123SNe_Vectors}
\caption{Illustration and definition of the chemical displacement for two example SNe. Combining the yields of two SNe with different progenitor masses results in an effective displacement of the ISM metal abundances. We define the chemical displacement as the resulting vector field of this operation.}
\label{fig:illudis}
\end{figure}

\begin{figure}
\centering
\includegraphics[width=0.47\textwidth]{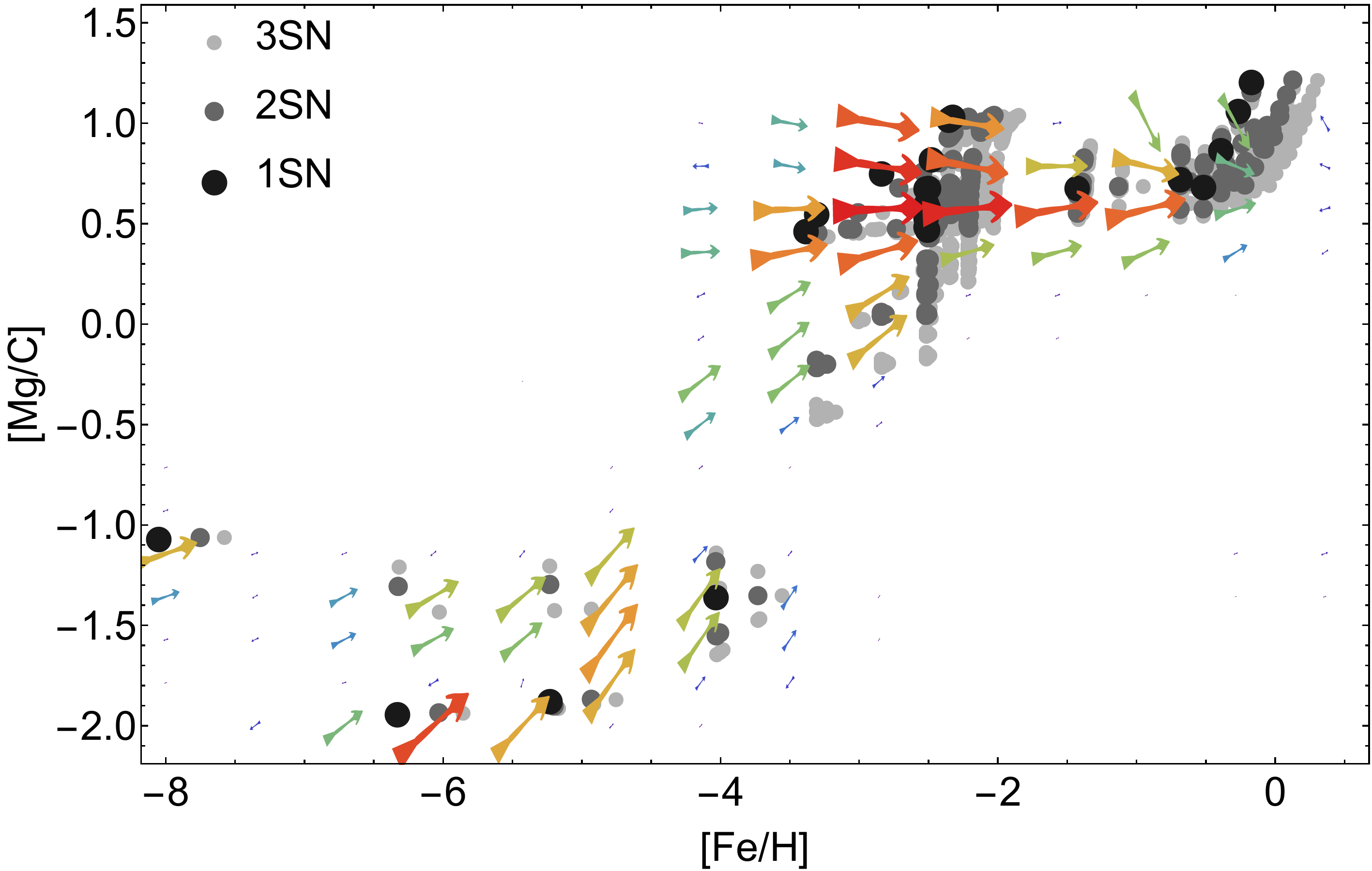}%{123SNe_Vectors_MgC}%{VectorField_Mg-over-C}
\caption{Illustration of the chemical displacement vector field of [Mg/C] and [Fe/H] for 25 SN progenitor masses, according to our fiducial IMF. The black points indicate the yields of single SNe for different progenitor masses. The $25^2$ dark grey points indicate the abundance ratios produced by combining the elemental yields of all possible combinations of two SNe from our set of 25. Similarly, the $25^3$ light grey points represent the combined yields of three Pop~III SNe. This plot illustrates how adding yields from several SNe changes the typically expected elemental ratios. For hydrogen, we assume a constant dilution mass of $7\times 10^5\Msun$, which is the median hydrogen mass in our sample of halos that are about to form second generation stars. The actual hydrogen mass and hence [Fe/H] might vary, but such an offset will not change the qualitative results for [Mg/C]. The length of the vectors illustrates the local magnitude of the chemical displacement field. The dynamical range of the vectors is decreased for better illustration and their length is therefore not to scale. The color of the arrows is an additional qualitative guidance to illustrate the magnitude of the vector field.}
\label{fig:streamLines}
\end{figure}

Our new diagnostic is based on the chemical displacement, which is illustrated in Fig.~\ref{fig:illudis}.
Commonly, the elemental abundances of observed EMP stars are plotted, but now we directly illustrate the SN yields and analyse how the chemical composition changes when we add the metal yields of two or more SNe.
%As illustrated in the figure, the chemical composition of the ISM depends on whether it is enriched by a faint SN, a core collapse SN, or both a faint and a core collapse SN.
Each possible combination of SNe yields defines two vectors which point to the resulting ISM abundance, as illustrated by the two arrows in this example. The resulting vector field of the successive mixing of SN yields from different progenitor stars defines the chemical displacement, which we show in Fig.~\ref{fig:streamLines}.
This vector field of the chemical displacement reflects changes in the abundances ratios when more than one SN contributes to the metal enrichment. The local magnitude of this vector field quantifies the tendency for enriched gas to be displaced from this region (i.e.\ to change its [Mg/C] and [Fe/H] abundances) when the elements of an additional SN are added.

To further quantify the chemical displacement, and the most promising elements for identifying mono-enriched second generation stars, we calculate the divergence of the chemical displacement field. The divergence describes the effective outward flux of a vector field that is emanating from a point. To guarantee numerical stability, we do not differentiate the resulting sparsely sampled vector field but apply Gauss' theorem: for each point where a displacement vector starts, we add the length of this vector to the divergence of this point. Where a displacement vector ends, we subtract the length of this vector from the divergence of this point.

Regions in abundance space with a high negative divergence attract SN yield contributions from other regions of the abundance space. Conversely, areas with a high positive divergence represent regions for which mixing with the yields of a second SN shifts the elemental abundances out of this region.

%TBD Since attracting regions with a high negative divergence imply a strong possibility of degeneracy of the Pop~III SN enrichment channels, . More precisely,
The information about the exact enrichment channel cannot be reconstructed uniquely for stars in areas with a negative divergence.
Therefore, the divergence of the chemical displacement simultaneously quantifies the information loss that occurs when combining several SN yields. A negative divergence corresponds to a high degeneracy.

% TBD
% Generally, regions of low iron abundance and low magnesium (low [Mg/C]) or low carbon abundance (high [Mg/C]) are favorable regions to find mono-enriched second generation stars. Conversely, we identify the range of $-1 \lesssim$[Mg/C]$\lesssim 0.5$, where it is very unlikely to find a metal-poor star that has been enriched by only one previous Pop~III SN because no yields of single SNe are in this region. Besides this exemplary study for magnesium and carbon, one can use all observed abundances for an EMP star to calculate the multi-dimensional divergence of the chemical displacement to quantify the probability for mono-enrichment. This method goes beyond the usually used diagnostic based on the metallicity and [C/Fe].

% %

\subsubsection{Comparison to semi-analytical model}
To highlight the strengths and weaknesses of our new diagnostic we compare it to the probabilities of a star being mono-enriched, as derived from our cosmological model.

\begin{figure}
\centering
\includegraphics[width=0.47\textwidth]{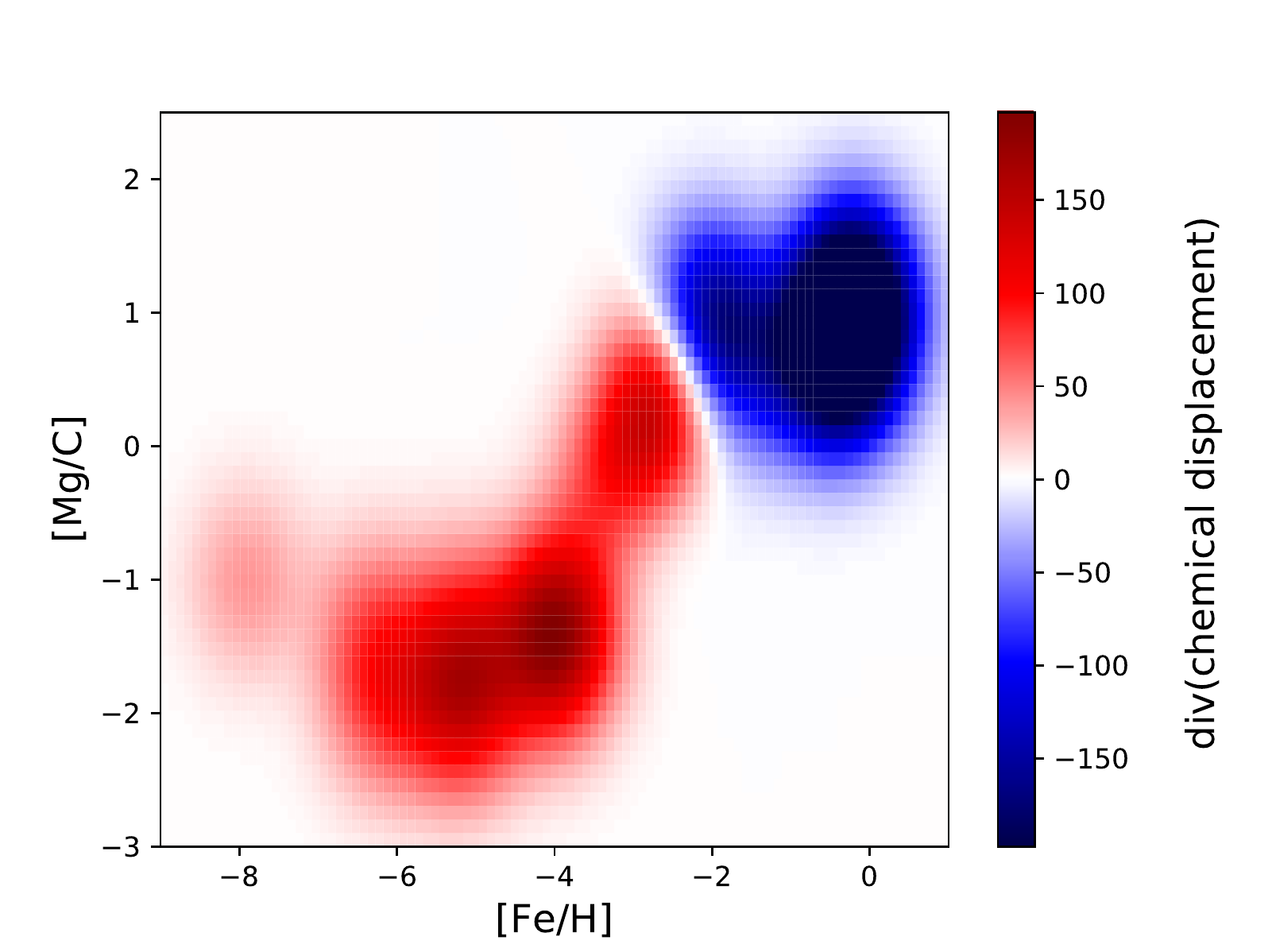}%{Mg-over-C_Map}
\caption{Divergence of the chemical displacement, based on the SN yields by \citet{nomoto13}. Positive values indicate promising regions to find mono-enriched second generation stars. Negative values represent attracting regions with a high chance of degeneracy due to yields being higher overall. To find mono-enriched second generation stars, EMP stars with [Mg/C] $<-0.5$ should be selected. Stars with [Mg/C] $\sim 1$ and [Fe/H]$\gtrsim -3$ are likely to be enriched by multiple SNe.}
\label{fig:div2D}
\end{figure}

The divergence map for [Mg/C] can be seen in Fig.~\ref{fig:div2D}.
This divergence map should be compared to Fig.~\ref{fig:pMgC} to see that we can reproduce the same trend with a more flexible method, fewer assumptions regarding the details of Pop~III star formation, and with less computational time. Our new diagnostic does not reproduce the high probability region at [Mg/C] $\gtrsim 0.5$ and [Fe/H]$< -5$ in the lower panel of Fig.~\ref{fig:pMgC} because this high probability of mono-enrichment emerges from only two events with a very small hydrogen dilution mass. The hydrogen dilution mass is assumed to be constant in our calculation of the chemical displacement.

The figure also highlights the dominating nature of core collapse ([Fe/H]$\, \sim -2$) and pair-instability SNe ([Fe/H]$\sim -1$), both around $0.5<$[Mg/C]$<1.0$. These SN have high yields of Mg, C, and Fe and therefore dominate the metal mass budget over those of other SNe, after they were combined with the metal yields of a second or third SNe. This illustrates that it is generally difficult with our diagnostic to uniquely identify mono-enriched second generation stars that have abundance ratios close to those produced by a SN with a high mass of ejected metals.

This implies an important consequence for EMP stars that formed from the gas enriched by such a dominating Pop~III progenitor. Since the dominating Pop~III SN has large absolute metal yields, it can thus not be excluded that another progenitor SNe with a lower yield is ``hidden'' in the observed stellar signature. We thus conclude that only EMP stars enriched by one (faint) SN with a small absolute metal yield can be clearly identified as mono-enriched stars.

\begin{figure}
\centering
\includegraphics[width=0.47\textwidth]{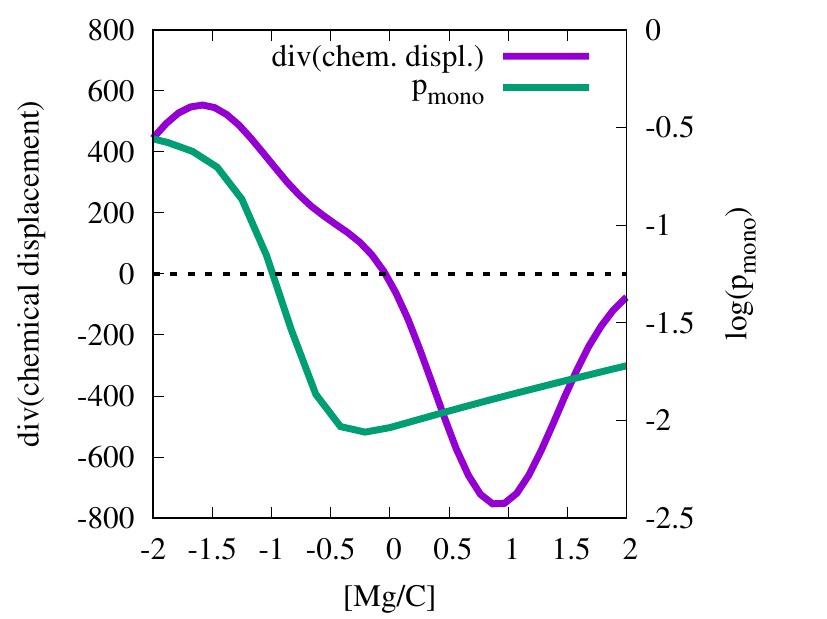}%{Compare_div}
\caption{Comparison of the divergence of the chemical displacement (purple, left y-axis) to the probability of being mono-enriched based on our cosmological model (green, right y-axis). Due to the different units, the two methods can only be compared at a qualitative level. The overall behaviour and predictive power of the two diagnostics is generally the same: the maximum of the divergence of the chemical displacements corresponds to  $p_\mathrm{mono} \gtrsim 30\%$ and a negative divergence to $p_\mathrm{mono} \lesssim 3\%$.}
\label{fig:compare}
\end{figure}

A direct comparison for [Mg/C] in one dimension is given in Fig.~\ref{fig:compare}.
The different units of the two diagnostics allows only a qualitative comparison. Both methods identify the range below [Mg/C] $\lesssim -1$ as a promising region to find mono-enriched second generation stars. The peak value of the divergence of the chemical displacements corresponds to probabilities around 30\% for finding mono-enriched stars using the results from our semi-analytical model.
%In the range $-0.5 \lesssim$ [Mg/C] $\lesssim 0$, we also find a positive divergence of the chemical displacement although the corresponding probability for mono-enrichment is $<1\%$.  A comparison with e.g.\ Fig.~\ref{fig:pMgC} shows that this region is poorly sampled and almost no second generation stars in our model fall in this region. Therefore, the probability to find mono-enriched second generation stars is low in this range. However, a second generation star in this region would have a high tendency to move away when the chemical yields that make its natal gas cloud would have been combined with the yields of another SN. Therefore, the divergence of the chemical displacement is positive in this region. Finally, the region around [Mg/C] $\sim 0.7$ shows a local peak of the probability to find mono-enriched second generation stars, despite the fact that the divergence of the chemical displacement is negative.
Conversely, a negative divergence of the chemical displacement at [Mg/C]$>0$ corresponds to probabilities of $\lesssim 3\%$ for mono-enrichment. This indicates an attracting region with a high degeneracy between mono- and multi-enriched second generation stars.

This comparison shows a qualitative agreement between our semi-analytical cosmological model and the new diagnostic based on the divergence of the chemical displacement. We highlight again that this new diagnostic is cheaper, more flexible and involves fewer free parameters than the full cosmological model.
%The overall probability to find mono-enriched second generation stars in this region is thus only $\lesssim 3\%$.

\subsubsection{Divergence in 1D}
In the previously presented example, this diagnostic tool was derived in 2D for two elemental abundance ratios but it can also be applied in higher-dimensional vector spaces if information is available on additional abundance ratios or for a single elemental ratio to obtain the trends with these elements.
%, with further constraining elemental ratios.% as we show in Fig.~\ref{fig:1Ddiv}.

\begin{figure}
\centering
\includegraphics[width=0.47\textwidth]{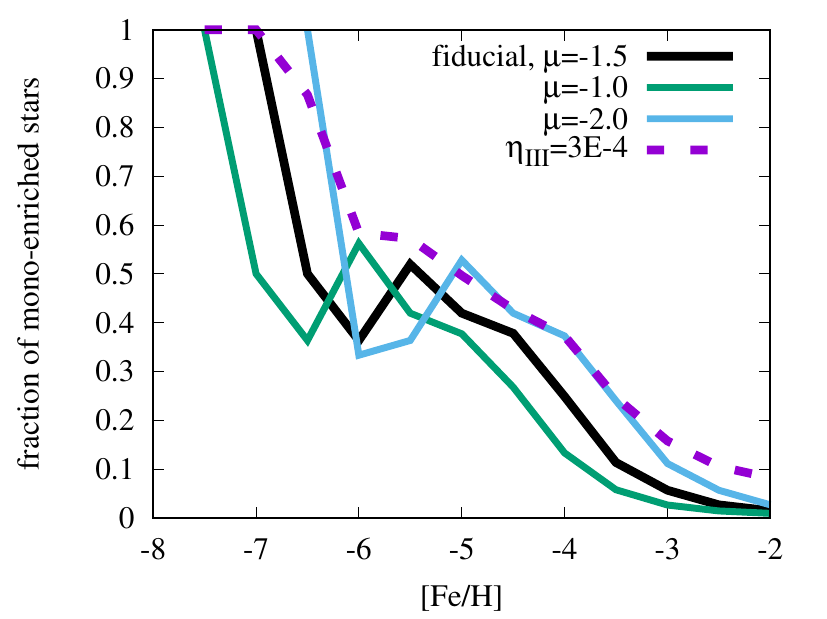}%mono-fraction
\caption{Fraction of mono-enriched second generation stars as a function of the metallicity, based on our semi-analytical model. In our fiducial model, this fraction is 100\% for [Fe/H]$\leq-7$ and around 40\% in the range $-6\lesssim$[Fe/H]$\lesssim-4$. There can also be multi-enriched second generation stars at [Fe/H]$\lesssim -6$, although the probability for this case is small. This distribution depends on the SFE, $\eta_\mathrm{III}$, and on the assumed fraction of hydrogen that mixes with the ejected metals after an enrichment event, $10^\mu$.}
\label{fig:monofrac}
\end{figure}

We canonically expect to find mono-enriched second generation stars at the lowest metallicities \citep{ryan96}, as we show in Fig.~\ref{fig:monofrac}.
This distribution is affected by the Pop~III SFE and by the efficiency of metal mixing. Allowing the ejected metals to mix on average with a larger fraction of the gas in a halo ($\mu=-1.0$) shifts this distribution to lower metallicities compared to the fiducial model. A lower SFE yields higher values for the fraction of mono-enriched second generation stars at all metallicities because we expect fewer Pop~III SN to explode per halo. The fraction of mono-enriched stars increases with decreasing metallicity. Therefore, the [Fe/H] values on the abscissae of figures~11-14 do not represent novel information as such.

\begin{figure}
\centering
\includegraphics[width=0.47\textwidth]{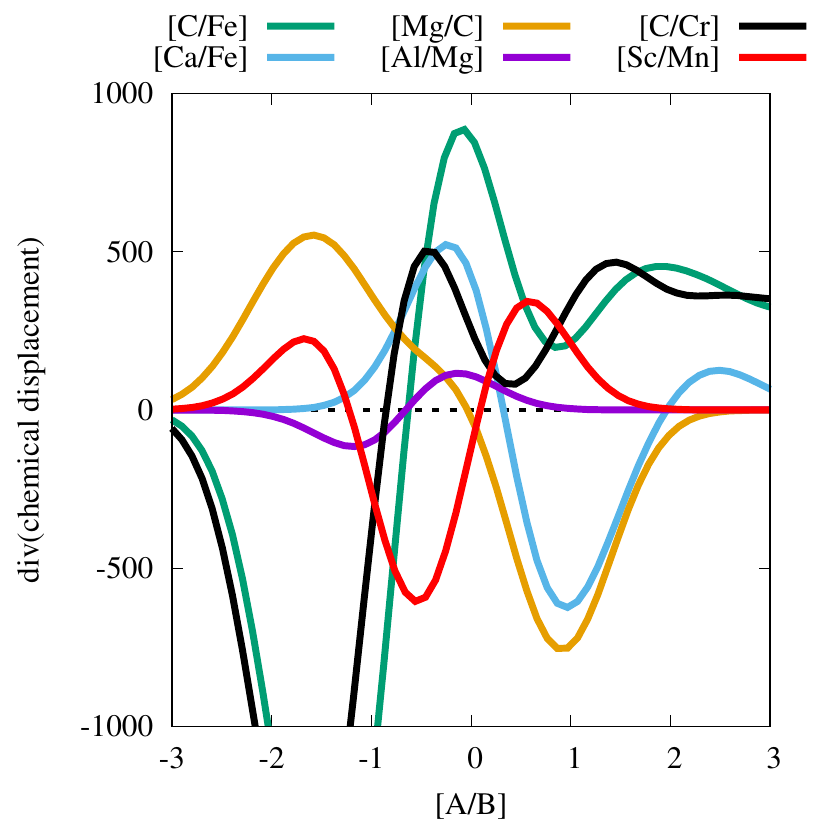}%{divChemDis}
\caption{Divergence of the chemical displacement for various elemental abundance ratios. For example, [Mg/C]$<-0$ and [C/Fe]$\gtrsim 0$ are promising diagnostics to find mono-enriched second generation stars. In contrast, [Al/Mg] cannot be used because the overall range of possible elemental ratios ($<1$\,dex) is of the same order as errors in the model and abundance estimates.}
\label{fig:1Ddiv}
\end{figure}

In a further step, we therefore calculate the 1D divergence of various elemental ratios as an additional diagnostic. The results are shown in Fig.~\ref{fig:1Ddiv}.
This not only highlights the most promising abundance ratios that should be used to find mono-enriched second generation stars, but it also allows us to compare different element diagnostics: the absolute value of the divergence quantifies how strongly a certain region is going to be attracting or repulsing. Moreover, it is important to examine the size of the difference in the abundance ratios between regions of positive and negative divergence. If this difference is too small, as for [Al/Mg], uncertainties in both the aluminium and magnesium yields weaken the predictive power. A reliable diagnostic requires a peak of high divergence that is significantly separated from regions with negative divergence. 

Although [C/Fe] $\sim -0.2$ seems to be a promising value to find mono-enriched stars, we note that this abundance range not only reflects the typical yield of Pop~III core-collapse SNe but, also corresponds to the yield from core-collapse SNe that arise from Pop~I metal-rich stars. This immediately illustrates the limitations of our diagnostic tool, as it does not include the yields of all possible metal sources self-consistently.

%Overall, the [Mg/C] ratio is a more reliable indicator than [C/Fe] when it comes to identifying second generation stars. The latter ratio is often large, not only for the faint Pop~III SNe yields but also when produced in $150\Msun$ PISN (see Fig.~\ref{fig:CFe}). However, these PISNe lead to regions with a low divergence, and hence, a high degeneracy. A high [C/Fe] could also point to the mixture of the yields of a $\sim 150\Msun$ PISN and a CCSN. On the contrary, low [Mg/C] is found only in faint Pop~III SNe yields, and is thus a more reliable indicator of mono-enriched second generation stars. We thus conclude that only EMP stars enriched by one (faint) SN with a small absolute metal yield can be clearly identified as mono-enriched stars.

\subsubsection{Applying the new diagnostic to EMP stars}
\begin{figure*}
\centering
\includegraphics[width=0.47\textwidth]{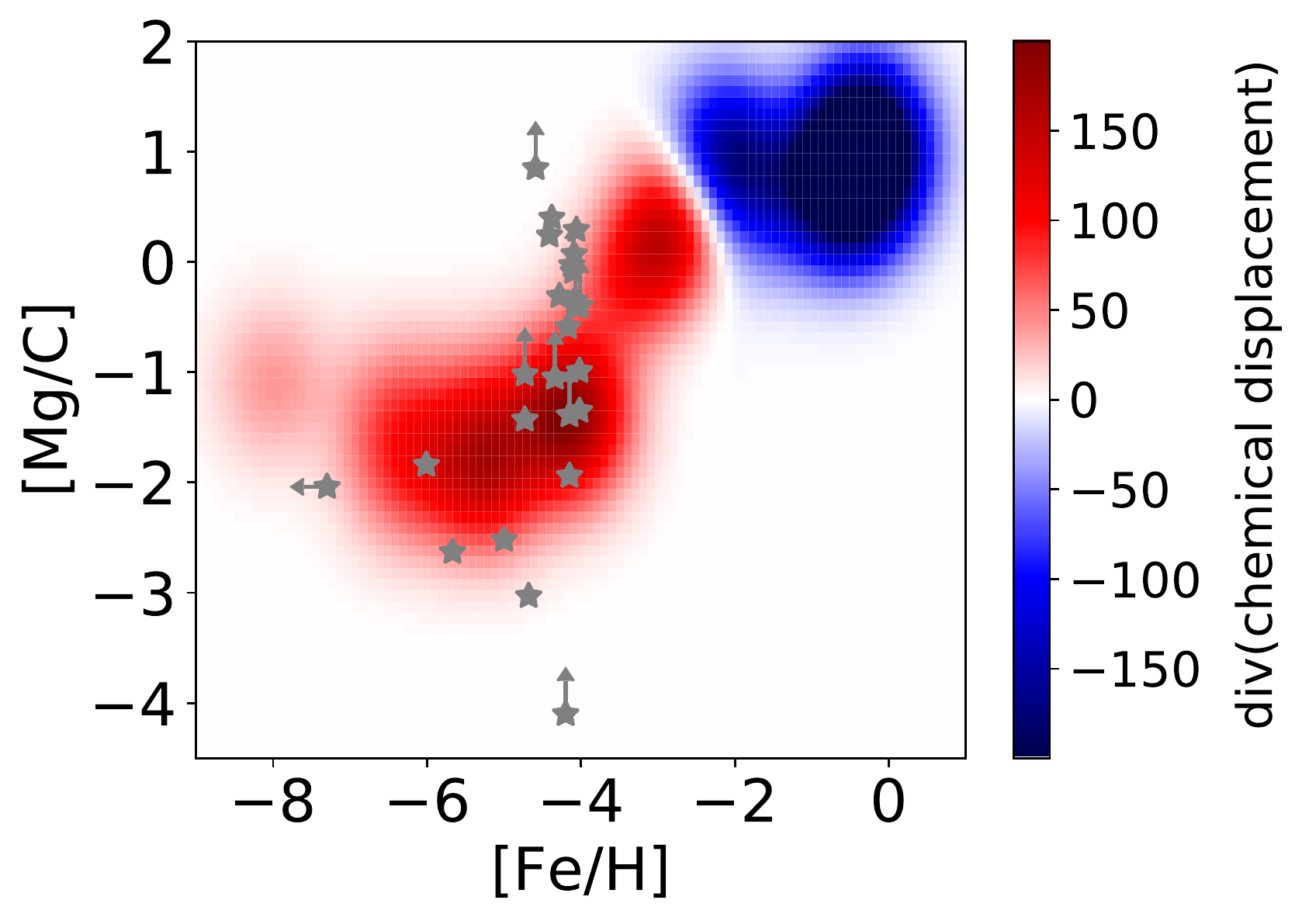} %{Mg-over-C_EMP_Map}
\includegraphics[width=0.47\textwidth]{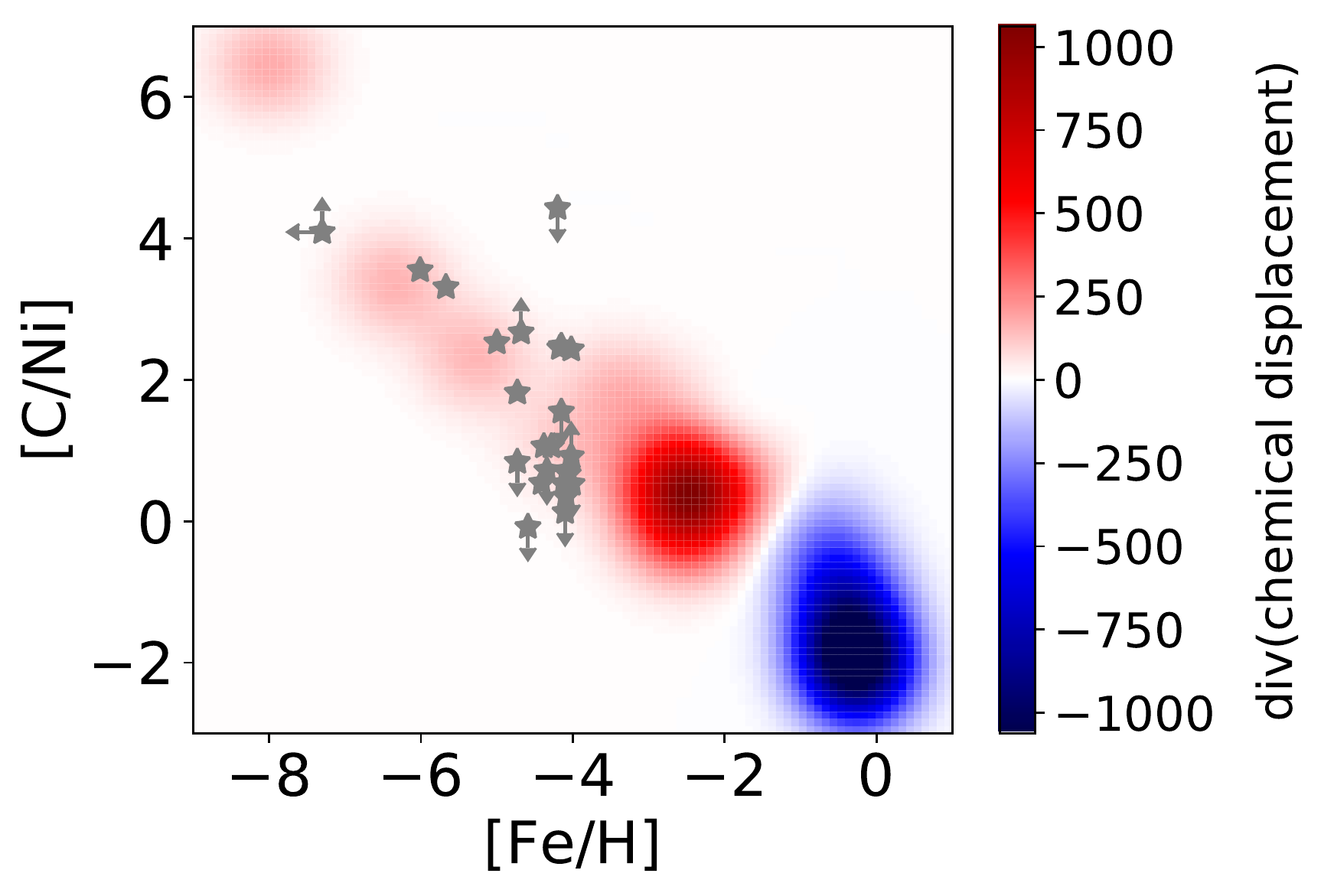} %{C-over-Ni_EMP_Map}
\includegraphics[width=0.47\textwidth]{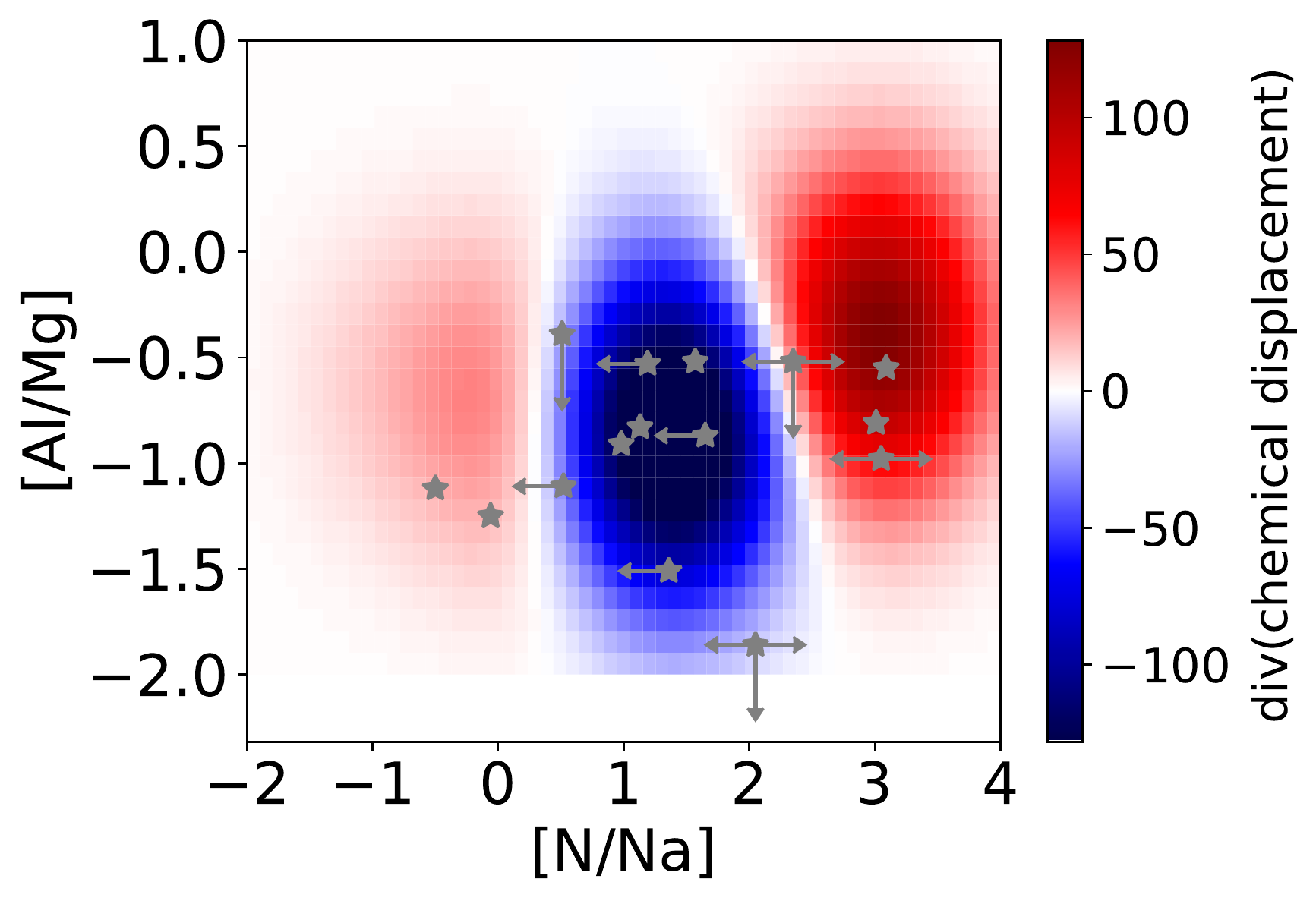} %{Mg-over-Ti_EMP_Map}
\includegraphics[width=0.47\textwidth]{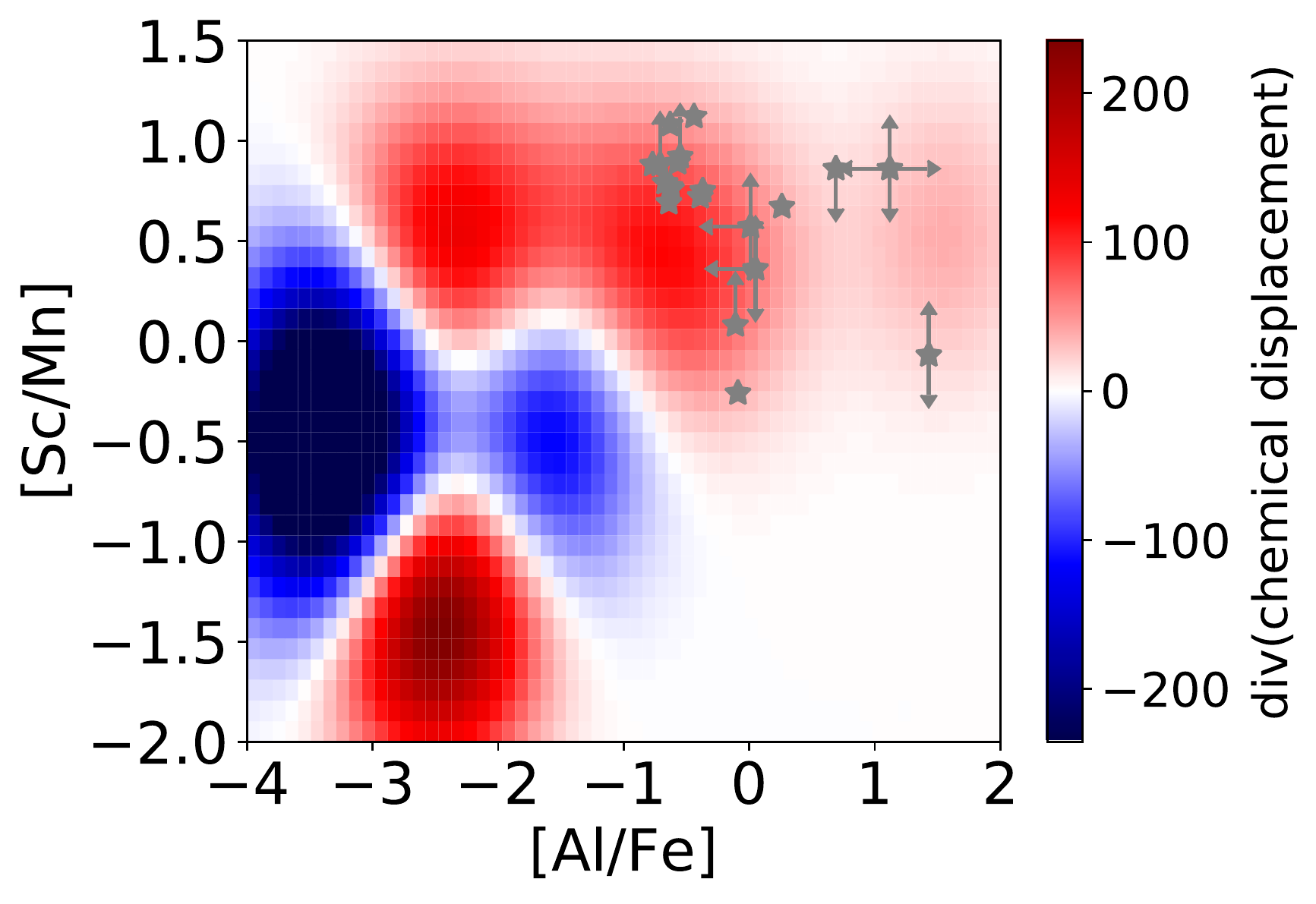} %{Sc-over-Mn_EMP_Map}
\caption{Maps of the divergence of the chemical displacement for different elemental ratios overplotted with a sample of EMP stars from the JINAbase \citep{abohalima17}. Upper limits on the measured abundances are illustrated as arrows. This representation allows us to infer the trends of the stars being mono- or multi-enriched. Stars in regions with a high positive divergence (red) are likely to be mono-enriched, whereas a high negative divergence (blue) indicates a possible degeneracy of elemental yields and therefore a high probability of being multi-enriched. These divergence maps are based on the SN yields by \citet{nomoto13}.}
\label{fig:EMPs}
\end{figure*}

We apply our new diagnostic to a selection of observed stars from the JINAbase \citep{abohalima17}. We select all stars with [Fe/H]$<-4.5$ and chose the elements for which this sample is most complete, see Fig.~\ref{fig:EMPs}.

We will provide a more detailed comparison to observations in a follow-up study but already briefly summarize the main conclusions and shortcomings here. Most of these EMP stars have a positive divergence of the chemical displacement and are therefore likely to be mono-enriched. The [N/Na] ratio of at least three stars, however, corresponds to a negative divergence: DC$-$38245 \citep{cayrel04}, CS30336$-$049 \citep{yong13}, and HE~1327$-$2326 \citep{frebel08}. Some other stars with only upper limits are in the same region of negative divergence, which could be interpreted as a signature for multi-enrichment. The yields of nitrogen and sodium are sensitive to stellar rotation, which is not included in the models by \citet{nomoto13}, and their abundance is difficult to derive accurately due to artificial nitrogen enhancement through the CNO cycle. Also differences between 1D and 3D non-local thermodynamic equilibrium stellar atmosphere models can affect the abundance by more than $0.5$\,dex.

%Several stars are always in or close to regions with a positive divergence, such as SMSSJ031300.36$-$670839.3 \citep{keller14}, SD1313$-$0019 \citep{frebel15b}, HE~2239$-$5019 \citep{hansen15}, and are therefore likely to be mono-enriched. Other stars have a negative divergence for [Mg/Ti] vs. [N/Na], but a positive divergence for the other diagnostic: HE~0107$-$5240 \citep{collet06} and HE~1327$-$2326 \citep{frebel08}. These cases require a careful analysis since a negative divergence also indicates regions of high degeneracy and is not a sufficient criterion for multi-enrichment.

Some observed EMP stars are outside the boundaries of our model, i.e.\ in regions for which we do not predict a value of the divergence. This is due to our limited sample of elemental yields, which for example does not include hypernovae or mass transfer from an asymptotic giant branch star across a binary system as possible sources of metals. Moreover, we only assume SN models with one explosion energy per progenitor mass and no rotation of the progenitor star. These effects would increase the diversity of possible elemental ratios and therefore widen the parameter space for which we can calculate the divergence of the chemical displacement. In a future study, we will improve this diagnostic, by taking into account a larger variety of sources for metals in the early Universe.

%hydrogen dilution?

\section{Discussion}
Our novel diagnostic based on the divergence of the chemical displacement can be applied to assess the likelihood of a star to be mono-enriched. A representation of the divergence such as in Fig.~\ref{fig:1Ddiv} or Fig.~\ref{fig:EMPs} will be most useful to classify metal-poor stars based on their measured abundances. However, the divergence of the chemical displacement cannot be directly translated into a probability of mono-enrichment. It rather reveals regions with a positive divergence in the multi-dimensional space of stellar abundances, which are dominantly populated by mono-enriched stars. A negative divergence is not a sufficient condition for multi-enrichment. A mono-enriched star can also be found in regions with a negative divergence, if it formed from gas enriched by an SN with high metal yields. In a future project we will improve this diagnostic and apply it to further EMP stars.

%We discuss limitations of this model below and will improve this approach in a follow-up study.

%This diagnostic fails in poorly sampled regions, where our model does not predict any second generation stars. 

%DLAs? \citep{kobayashi11}: their yields are in agreement with nucleosynthetic yields of faint SNe and not consistent with PISNe yields. Also \citet{cooke11,cooke17}.

\citet{ji15} also considered how the abundance of second-generation stars would be affected by forming a small multiple of Pop III stars in minihalos. They focussed on two specific scenarios of second-generation star formation: immediate gas recollapse in a minihalo and delayed formation in atomic cooling halos. These cases are applicable in the earliest stages of Pop~III and Pop~II star formation, but at later times global radiative feedback becomes important. Our model includes the effect of external radiation in a cosmological context, extending its applicability to lower redshifts. \citet{ji15} also focussed on specific element ratios with critical ratios to investigate the carbon-enhanced and PISN signatures. Our new chemical divergence formalism generalizes their approach and allows more efficient searching of the ideal ratios in the full abundance space, independent of the specific assumptions of star formation.

%Compare to Renaissance simulation (e.g. SFRd)

\subsection{Which element traces the total metal content best?}
\begin{figure}
\centering
\includegraphics[width=0.47\textwidth]{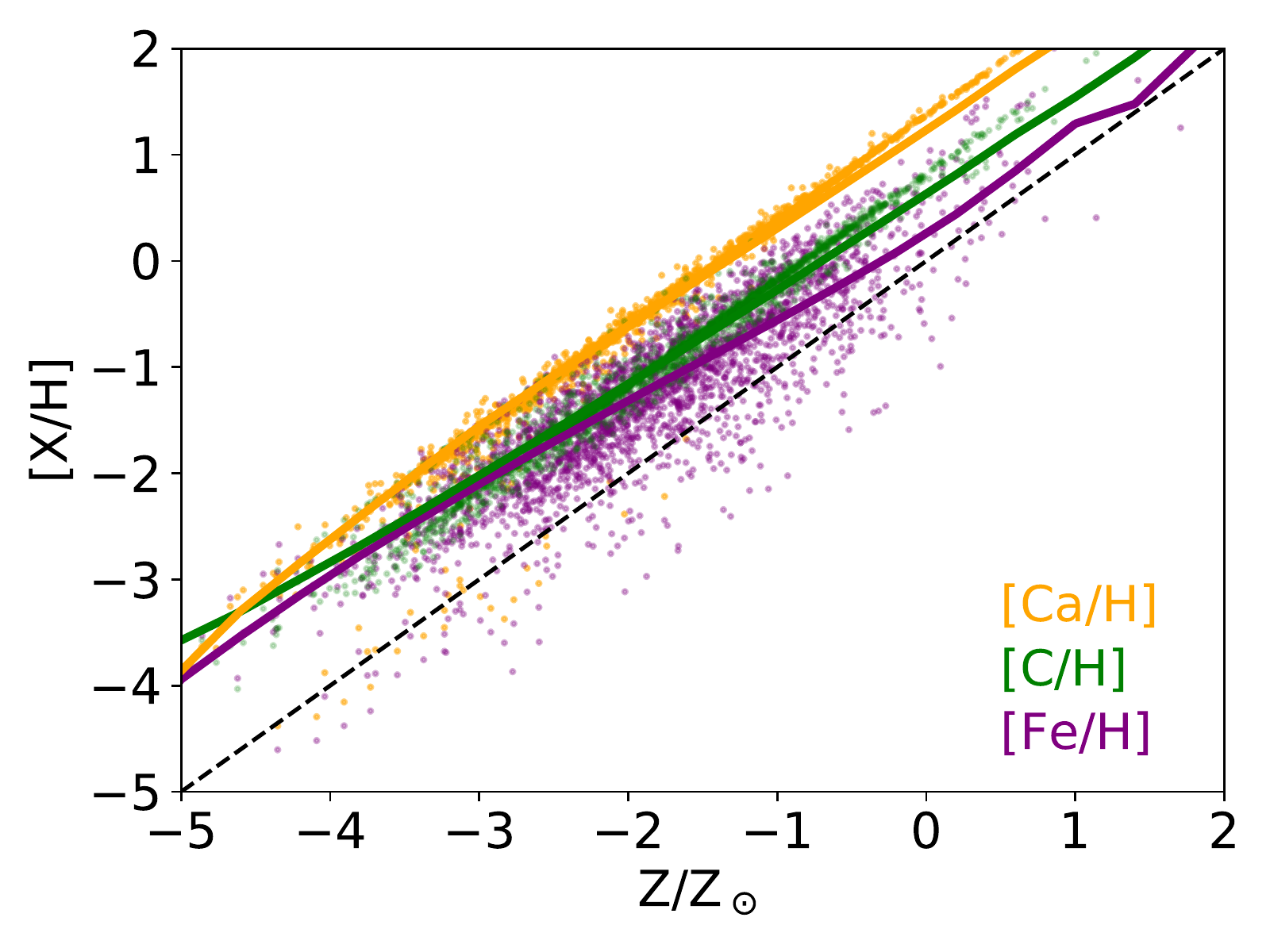}%{fiducial_TracerZ}
\caption{Individual elemental abundances as a function of the total metal content of second generation stars in our model. The three coloured lines indicate the binned medians of these distributions. Iron and carbon are better tracers of the total metallicity of second generation stars than calcium. Whereas the [Fe/H] distribution lies closer to the diagonal (black dashed) and that for [C/H] slightly above, the mostly likely value for [Ca/H] is one dex above the corresponding Z/Z$_\odot$. For better illustration, we only plot a small subset of all second generation stars.}
\label{fig:Z}
\end{figure}

In theoretical models of cosmic chemical evolution or of the formation of the first low-mass stars, the total metal content is of fundamental importance. The metal content of a star is defined as the relative abundance of all elements heavier than helium, relative to our Sun, which consists of $\sim2$\% metals: Z/Z$_\odot=\mathrm{log}_{10}(M_\mathrm{metals}/(0.02M_\mathrm{gas}))$. To connect this total metallicity to the observed abundances of individual elements, we show in Fig.~\ref{fig:Z} which element is a reliable tracer for the total metal content of a star.

We find that [Ca/H] is on average about one dex above Z/Z$_\odot$ for second generation stars, albeit with a large scatter. Iron and carbon abundances are more reliable tracers for the total metal content of a star and the usage of calcium can lead to severe misinterpretations: stars with an estimated [Ca/H]$\approx -2$ may have an overall metallicity of Z/Z$_\odot <-3$ and therefore be falsely rejected for any spectroscopic follow-up study. These results are a consequence of our assumed SN yields, which also show an IMF-averaged offset of $\sim0.5$\,dex for [Ca/C] and [Ca/Fe].
We also find the inverse possibility, mostly for calcium and iron, that a star with a low individual abundance of these elements can still have a high total metal content (below dashed diagonal). These results are insensitive to the treatment of hydrogen dilution because all derived ratios scale equally with the hydrogen mass.

We further quantify the scatter of the distributions and find that carbon at Z/Z$_\odot <-3$ has the smallest standard deviation of $\sim 1$\,dex and both calcium and iron have a scatter of $\sim 1.2$\,dex in the same range of the overall stellar metal content.
%Other elements with high absolute yields, such as oxygen or chlorine, also correlate well with the total metal content, but are more difficult to measure spectroscopically.
%The trend that carbon traces best the overall stellar metal content with the smallest scatter is a consequence of the high fraction of CEMP stars at low metallicities: for these stars, carbon is a major element in absolute mass, therefore also dominates the overall stellar metallicity, and consequently reproduces this relation best.

The commonly used pre-selection method is to identify EMP candidates based on the Ca K line \citep[see e.g.][]{keller07,koch16,starkenburg17} because it is strong, easy to see in low-quality data, and therefore most efficient regarding telescope time. Metal-poor stars show weaker calcium absorption features than more metal-rich stars. The additional use of a carbon-sensitive filter (e.g.\ the G-band around $\sim$4300\,\AA) could yield a more reliable photometric estimate for the total metal content, although this suggestion has to be treated with caution when explicitly targeting CEMP stars.

\begin{figure}
\centering
\includegraphics[width=0.47\textwidth]{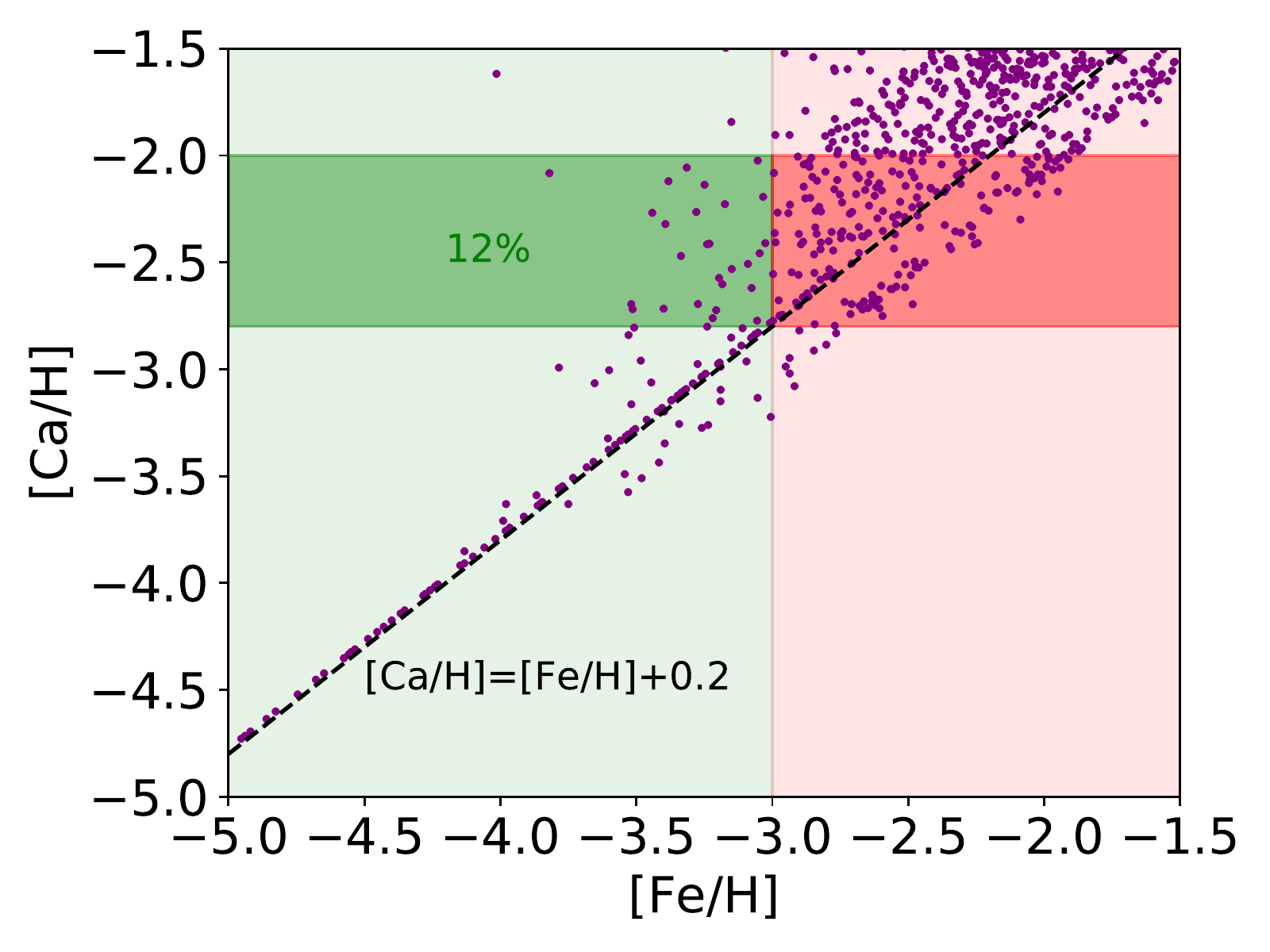}%{TracerZ_CaFe}
\caption{Calcium against iron abundance for second generation stars. The dark region represents the fraction of EMP stars ([Fe/H]$<-3$) that a survey would miss if it only selects stars with [Ca/H]$<-2.8$ instead of [Ca/H]$<-2.0$ for a higher resolution follow-up observation.}
\label{fig:CaFe}
\end{figure}

We further analyse the fraction of EMP stars that a survey misses due to a too conservative calcium-based pre-selection. In Fig.~\ref{fig:CaFe} we show an example model for the fraction of falsely rejected stars.
If we are interested in EMP stars with [Fe/H]$<-3$ and assume that calcium traces iron with [Ca/H]$=$[Fe/H]$+0.2$, a survey would reject stars with [Ca/H]$>-2.8$ as too metal-rich. However, in the range 
\begin{equation}
-2.8<\mathrm{[Ca/H]}<-2.0
\end{equation}
we find $12$\% of second generation stars with [Fe/H]$<-3$. In particular, PISNe with a progenitor mass around $\sim 150\Msun$ eject material with high [Ca/Fe] yields \citep{karlsson08} and so second generation stars enriched primarily by these PISNe will have high [Ca/Fe]. Our estimate can be used as an approximation for the completeness of surveys, although we note that our simulated sample might not be complete in this calcium range since we do not include enrichment by later generations of star formation. For an assumed relation of [Ca/H]$=$[Fe/H]$+0.4$, we still find $\sim 11$\% of EMP stars in the corresponding range \begin{equation}
-2.6<\mathrm{[Ca/H]}<-2.0.
\end{equation}

\subsection{Caveats}
Our diagnostic and predictions based on the divergence of the chemical displacement are only as good as the underlying models for the SN nucleosynthetic yields. We use the tabulated SN yields as a function of the Pop~III progenitor mass by \citet{nomoto13} with additional models for faint SNe by \citet{ishigaki14}. In a future study we will improve our model by including the metal contributions from other enrichment channels, such as neutron star mergers, hypernovae, AGB stars, and Type Ia SNe. Moreover we will assess the sensitivity of our model to the assumed Pop~III SN yield models and derive a diagnostic based on elements that are least sensitive to the underlying model assumptions.

We include Pop~III star formation as a sub-grid model based on the random sampling of individual stars from a given IMF. However, UV feedback by the primary formed massive star in a minihalo might prevent the formation of further massive stars \citep{susa14,hosokawa16}. Such a suppression of further Pop~III stars with higher masses might result in a steeper slope of the IMF at higher masses. Moreover, we have no information on the exact position of the first stars in a minihalo. Therefore we cannot take into account the effect of SNe that explode off-centre in the halo and have different metal ejection fractions, mixing efficiencies, or recovery times for the ISM.

Throughout the paper we do not track individual second generation stars. We rather follow their formation events and assume that such a burst of star formation creates a chemically homogeneous population of second generation stars. Therefore we cannot make reliable predictions about the absolute number of second generation stars in our model. Moreover, the number of stars per halo might differ, depending on the environment and the available gas mass. Larger systems, which are more likely to experience multiple previous SNe, will also host more second generation stars. This is an additional bias that reduces the relative number of mono-enriched second generation stars, which tend to form in less massive systems.

\section{Summary and Conclusion}
EMP stars in the MW provide a unique way to probe the mass distribution of the first stars. They carry the characteristic chemical fingerprint of the SN that enriched the gas from which they formed. A comparison of their observed chemical abundances with models of Pop~III SNe allows us to determine the Pop~III progenitor masses of the SNe. To fully exploit this method and avoid degeneracies in the fitting of the SN yields, it has to be applied to mono-enriched second generation stars.

In this paper, we have presented a novel diagnostic to identify this precious subclass of mono-enriched stars. We model the first generations of star formation with a semi-analytical model, based on dark matter merger trees from the Caterpillar simulations \citep{griffen16}. We find that the Pop~III star formation efficiency, the primordial IMF, the mixing efficiency of metals with the ISM, and the fraction of faint SNe are the main parameters to calibrate our model. The MDF and fraction of CEMP stars as observational constraints are best reproduced by our fiducial model with a logarithmically flat IMF in the mass range $3-150\Msun$. With a two-sample KS test we can exclude a Pop~III IMF that extends up to $M_\mathrm{max}=300\Msun$ at the 95\% level. In our model, PISNe from stars with masses of $\gtrsim 200\Msun$ fail to reproduce the MDF at [Fe/H]$\leq -3$ due to their high absolute metal yields.

Mono-enriched stars account for only $\sim 1\%$ of second generation stars in our fiducial model. This fraction is a strong function of the primordial SFE and we provide an analytical formula to independently calculate this fraction for different model assumptions (Eq. \ref{eq:pmono}). Dwarf satellites have the highest stellar fraction of mono-enriched second generation stars because they formed the majority of their stellar population early on. Satellites with $M_h<10^9\Msun$ host $10-100\%$ second generation stars and satellites with $M_h \lesssim 10^8\Msun$ contain only second generation stars, some of them only mono-enriched second generation stars. The specific numbers have to be treated with caution, since they are affected by uncertainties in the abundance matching.

We have also presented a novel analytical diagnostic to identify mono-enriched stars, based on the divergence of the chemical displacement. This new diagnostic allows to derive the likelihood of mono-enrichment independently from most parameters that govern the first billion years. The fraction of mono-enriched second generation stars is 100\% for [Fe/H]$\leq-7$ and around 40\% in the range $-6\lesssim$[Fe/H]$\lesssim-4$. We also present additional elemental ratios that are reliable tracers for mono-enrichment, such as [Mg/C]$<-1$, [Sc/Mn]$<0.5$, [C/Cr]$>0.5$, or [Ca/Fe]$>2$.
%We find that high and low ratios of [Mg/C] are a promising diagnostic to identify mono-enriched second generation stars.
%Our results can be applied as selection criteria for surveys and therefore contribute to a deeper understanding of EMP stars and their progenitors.

The chemical imprint of SNe with little ejected metals could be hidden in the abundance patterns from stars with more metals and consequently only faint SNe can be uniquely identified as being mono-enriched. Thus, focussing on mono-enriched stars biases the interpretation towards Pop~III progenitors with low metal yields. A negative divergence of the chemical displacement does not mean that such a star is multi-enriched, but that there is a certain possibility that this abundance pattern is the result of multi-enrichment.

The results of our study provide powerful diagnostic to interpret extensive photometric and spectroscopic data of metal-poor stars in the MW and its satellites, which will be available in the next decades. 
Specifically, our findings can be applied to data from narrow-band photometric surveys covering the Ca H\&K feature, such as the Pristine survey \citep{starkenburg17}, which provide unbiased view of the most metal-poor stars up to a large Galactic distance.
Next-generation low and medium-resolution spectroscopic facilities such as WEAVE \citep{bonifacio16}, or PFS \citep{takada14} are suitable for directly identifying mono-enriched second generation stars among MW field halo and dwarf satellite's stars, which will be the best targets for a follow-up high-resolution spectroscopy.
High-resolution spectroscopic surveys for large samples of stars in the MW are also on-going or planned in the near future with high-resolution multi-object spectrographs such as the APOGEE \citep{majewski17}, GALAH \citep{deSilva15}, and the 4MOST \citep{feltzing17} projects. 

\subsection*{Acknowledgements}
We thank the reviewer for constructive suggestions and careful reading of the manuscript. The authors would like to thank Ken'ichi Nomoto, Gen Chiaki, Hajime Susa, Kazu Omukai, and Wako Aoki for valuable discussions and helpful contributions. TH is a JSPS International Research Fellow.
%TH acknowledges funding under the European Community's Seventh Framework Programme (FP7/2007-2013) via the European Research Council Grants `BLACK' under the project number 614199.
The authors were supported by the European Research Council under the European Community's Seventh Framework Programme (FP7/2007 - 2013) via the ERC Advanced Grant `STARLIGHT: Formation of the First Stars’ under the project number 339177 (RSK) and via the ERC Grant `BLACK' under the project number 614199 (TH). SCOG and RSK also acknowledge funding from the Deutsche Forschungsgemeinschaft via SFB 881 `The Milky Way System' (subprojects B1, B2, and B8) and SPP 1573 `Physics of the Interstellar Medium' (grant numbers KL 1358/18.1, KL 1358/19.2 and GL 668/2-1). AF is supported by NSF CAREER grant AST-1255160. This research has been supported by the Canon Research Foundation. This research is further supported in part by the National Science Foundation (NSF; USA) under grant No. PHY-1430152 (JINA Center for the Evolution of the Elements).  APJ is supported by NASA through Hubble Fellowship grant HST-HF2-51393.001 awarded by the Space Telescope Science Institute, which is operated by the Association of Universities for Research in Astronomy, Inc., for NASA, under contract NAS5-26555.  BWO was supported by the National Aeronautics and Space Administration (NASA) through grant NNX15AP39G and Hubble Theory Grant HST-AR-13261.01-A, and by the NSF through grant AST-1514700. Computational support for the Caterpillar simulations was provided by XSEDE through the grants (TG-AST120022, TG-AST110038). Computations were carried out on the compute cluster of the Astrophysics Division which was built with support from the Kavli Investment Fund, administered by the MIT Kavli Institute for Astrophysics and Space Research. This work was supported by World Premier International Research Center Initiative (WPI Initiative), MEXT, Japan.
%We further acknowledge support of the compute cluster of the MIT Astrophysics Division, which was built with support from the Kavli Investment Fund administered by the MIT Kavli Institute for Astrophysics and Space Research.

\bibliographystyle{mn2e}
\bibliography{2ndGenerationStars}
%\end{thebibliography}

%\bsp
\label{lastpage}

\end{document}